\documentclass[11pt]{article}

\usepackage[final]{acl}

\usepackage{times}
\usepackage{latexsym}

\usepackage[T1]{fontenc}

\usepackage[utf8]{inputenc}

\usepackage{microtype}

\usepackage{inconsolata}

\usepackage{graphicx}

\usepackage{algorithm}
\usepackage{algorithmic}

\usepackage{booktabs}
\usepackage{amssymb}
\usepackage{amsmath}

\usepackage{amsfonts}
\usepackage{textcomp}
\usepackage{xcolor}
\usepackage{colortbl}

\definecolor{Gray}{gray}{0.9}

\usepackage{multirow}
\usepackage{subcaption}
\usepackage{arydshln}

\usepackage{newfloat}
\usepackage{listings}

\definecolor{Gray}{gray}{0.9}

\usepackage{soul}

\definecolor{lightgray}{rgb}{0.95,0.95,0.95}
\definecolor{Gray}{gray}{0.9}
\definecolor{mygreen}{rgb}{0.0, 0.5, 0.0}
\definecolor{myred}{rgb}{0.8, 0.25, 0.33}
\definecolor{myblue}{rgb}{0.19, 0.55, 0.91}
\definecolor{uclablue}{rgb}{0.15, 0.45, 0.68}
\definecolor{ucladblue}{rgb}{0.0, 0.33, 0.53}
\definecolor{ucladdblue}{rgb}{0.0, 0.23, 0.36}
\definecolor{uclagold}{rgb}{1.0, 0.82, 0.0}
\definecolor{ucladgold}{rgb}{1.0, 0.78, 0.17}
\definecolor{ucladdgold}{rgb}{1.0, 0.72, 0.11}
\definecolor{boxgreen}{rgb}{0.02, 0.66, 0.02}
\definecolor{boxred}{rgb}{0.66, 0.1, 0.1}
\definecolor{boxblue}{rgb}{0.01, 0.01, 0.73}
\definecolor{mypink}{rgb}{.99,.91,.95}
\sethlcolor{lightgray}

\title{MEIC-DT: \underline{M}emory-\underline{E}fficient \underline{I}ncremental \underline{C}lustering for Long-Text Coreference Resolution with \underline{D}ual-\underline{T}hreshold Constraints}

\author{\textbf{Kangyang Luo$^{\spadesuit}$, Shuzheng Si$^{\spadesuit}$, Yuzhuo Bai$^{\spadesuit}$,  Cheng Gao$^\spadesuit$, Zhitong Wang$^{\spadesuit}$}\\
\textbf{Cheng Huang$^\spadesuit$, Yingli Shen$^\spadesuit$, Yufeng Han$^{\spadesuit}$, Wenhao Li$^\spadesuit$, Cunliang Kong$^\spadesuit$} \\ 
\textbf{Maosong Sun\thanks{Corresponding author}$^{\spadesuit}$$^\clubsuit$$^\bigstar$ }\\
$^{\spadesuit}$Department of Computer Science and Technology, Tsinghua University 
\\ $^\clubsuit$Institute for AI, Tsinghua University \\ $^\bigstar$Jiangsu Collaborative Innovation Center for Language Ability}

\begin{document}
\maketitle
\begin{abstract}
In the era of large language models (LLMs), supervised neural methods remain the state-of-the-art (SOTA) for Coreference Resolution. Yet, their full potential is underexplored, particularly in incremental clustering, which faces the critical challenge of balancing efficiency with performance for long texts. To address the limitation, we propose \textbf{MEIC-DT}, a novel dual-threshold, memory-efficient incremental clustering approach based on a lightweight Transformer. MEIC-DT features a dual-threshold constraint mechanism designed to precisely control the Transformer's input scale within a predefined memory budget. This mechanism incorporates two key components: a Statistics-Aware Eviction Strategy (\textbf{SAES}) and an Internal Regularization Policy (\textbf{IRP}). The SAES utilizes distinct statistical profiles from the training and inference phases for intelligent cache management. The IRP strategically condenses clusters by selecting the most representative mentions, thereby preserving semantic integrity. Extensive experiments on common benchmarks demonstrate that MEIC-DT achieves highly competitive coreference performance under stringent memory constraints. 
\end{abstract}

\section{Introduction}
Coreference Resolution~(CR) is a fundamental task in Natural Language Processing~(NLP), which aims to identify different text spans (i.e., mentions) within a document that refer to the same entity~\cite{karttunen1976discourse, lee2017end, lee2018higher}. It plays a critical role in a variety of downstream applications, such as text summarization~\cite{liu2024bridging}, knowledge graph construction~\cite{pan2024unifying, chen2025large}, question answering\cite{pan2024unifying, jang-etal-2025-ambiguity}, and named entity recognition~\cite{shang2025local}.
Current mainstream CR methods can fall into three types: \textit{supervised neural methods}~\cite{xu-choi-2020-revealing, wu-etal-2020-corefqa, kirstain2021coreference, lai2022end, otmazgin2022f, otmazgin2022lingmess, guo-etal-2023-dual, martinelli-etal-2024-maverick}, \textit{generative methods}~\cite{liu-etal-2022-autoregressive, bohnet-etal-2023-coreference, zhang-etal-2023-seq2seq}, and \textit{large language model-based methods}~\cite{le2023large, gan-etal-2024-assessing, zhu-etal-2025-llmlink}. Among these, supervised neural methods demonstrate significant advantages in terms of training cost, inference efficiency, while achieving competitive performance. However, their potential, particularly for challenging scenarios like long-text processing, has not yet been fully explored. 

\begin{figure*}[t]
  \centering
  \includegraphics[width=1.0\linewidth]{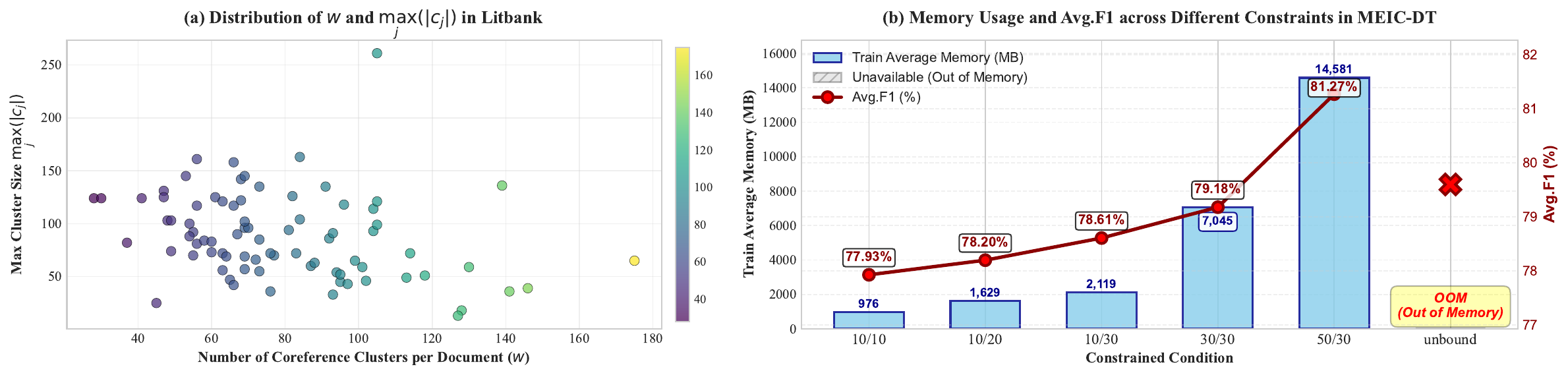}
  \caption{Analysis of motivations on the LitBank training corpus. Notably, the ``unbound'' condition failed due to out-of-memory (OOM) errors under our memory constraints.}
  \label{sta_onto_lit_fig:}
\end{figure*}

Within \textit{supervised neural methods}, incremental clustering---a strategy inspired by human incremental cognitive processing~\cite{altmann1988interaction}---is widely adopted for long-text scenarios. However, despite significant progress in CR, existing methods~\cite{xia-etal-2020-incremental, toshniwal-etal-2020-learning, guo-etal-2023-dual, martinelli-etal-2024-maverick} still face a critical challenge in balancing efficiency and performance during mention clustering. For example, conventional incremental clustering approaches~\cite{xia-etal-2020-incremental, toshniwal-etal-2020-learning, guo-etal-2023-dual} typically employ a fixed-size cache and a linear classifier to associate candidate mentions with coreference clusters (each represented as a single vector) tracked by the cache. These approaches decide whether a mention should join an existing cluster or form a new one. Upon reaching capacity, a predefined rule-based eviction strategy---LRU (\textit{Least Recently Used}) or a dual-cache mechanism combining LRU and LFU (\textit{Least Frequently Used})---removes certain entities to update the cache dynamically.  While computationally efficient, such approaches often fail to capture the complex nonlinear dependencies between mentions and clusters, and result in limited semantic representations due to excessive intra-cluster compression. Moreover, these strategies either rely on a single heuristic (e.g., LRU) that may discard essential clusters, or employ a dual-cache mechanism that adds to the complexity of cache management. Crucially, the same eviction strategy is used across both training and inference. This does not account for the inherent differences in information availability between the two phases and thus hinders the model's overall effectiveness.

Recently, \citet{martinelli-etal-2024-maverick} introduced a lightweight Transformer model and achieved SOTA performance. 
Its input is formed by concatenating the representation of a candidate mention with representations of all mentions from historical coreference clusters, effectively modeling the global intra-cluster structure and significantly improving clustering accuracy.
However, a major drawback is that the input size grows polynomially with the number of clusters and the number of mentions per cluster, causing a surge in memory usage and computational latency. This severely limits the model's practicality and scalability under typical academic computational resources.
\textbf{\textit{These considerations motivate our investigation into pushing the boundaries of supervised neural models in the era of LLMs.}}

To tackle the above issues, we propose \textbf{MEIC-DT}, a novel memory-efficient incremental clustering approach with dual-threshold constraints based on a lightweight Transformer, which aims to achieve competitive coreference resolution performance while adhering to a strict, predefined memory budget.
MEIC-DT formulates a \textbf{\textit{dual-threshold}} \textbf{\textit{constraint}} \textbf{\textit{mechanism}} that regulates the Transformer’s input scale along two key dimensions: first, by leveraging distinctive information available during training and inference, we design a \textbf{\textit{statistics-aware eviction strategy}} to preserve crucial coreference clusters within the constrained cache space; second, we introduce an \textbf{\textit{internal regularization policy}} that limits the number of mentions per cluster, enabling the selection of only the most representative mentions for computation under tight memory conditions.
To sum up, we highlight our contributions as follows:
\begin{itemize}
    \item  We propose a novel approach MEIC-DT, which reduces the memory and computational burden of Transformer-based coreference clustering while maintaining competitive performance. 
    \item We design a dual-threshold constraint mechanism, integrating a statistics-aware eviction strategy for phase-adaptive cache management and an internal regularization policy for cluster condensation, to ensure robust and efficient operation under stringent memory limits.
    \item Extensive experiments on several commonly used CR benchmarks~(i.e., OntoNotes, LitBank and WikiCoref) show that MEIC-DT achieves competitive coreference performance in memory-constrained scenarios. 
\end{itemize}

\begin{figure*}[t]
  \centering
  \includegraphics[width=1.0\linewidth]{Framework.pdf}
  \caption{The MEIC-DT Coreference Resolution pipeline. 
  The core innovation is a dual-threshold constraint mechanism---comprising a \textit{statistics-aware eviction strategy} and an \textit{internal regularization policy}---that enables high-performance, memory-efficient incremental clustering.
  }
  \label{Framework_fig:}
\end{figure*}

\section{Preliminaries}
\textbf{Notations.} 
Given a long text $D = [t_1, t_2, \dots, t_n]$ composed of $n$ tokens, let $\mathbf{H}=[\mathbf{h}_1, \cdots, \mathbf{h}_n] \in \mathbb{R}^{n\times d_h}$ denote the hidden representation of $D$ from a text encoder, where $d_h$ is the hidden dimension. 
During mention clustering, the set of candidate mentions output by a mention extractor is defined as $M = [m_1,m_2, \cdots, m_r]$, where each mention $m_i$ is represented by the start and end tokens $(t_{i,s}, t_{i,e})$. Historical coreference clusters (i.e.,  entities) are denoted as $\{c_j\}_{j=1}^w = \{[m_{j,1}^f, \cdots, m_{j,|c_j|}^g ]\}_{j=1}^w$, where $f$ and $g$ are indices of mentions in $M$ with $f \leq g$, and $w$ is the total number of clusters.

\textbf{Transformer-based incremental model.} To our knowledge, \citet{martinelli-etal-2024-maverick} were the first to employ a lightweight Transformer for mention clustering. 
Their model achieves SOTA performance on long texts by computing clustering probabilities between candidate mentions and existing clusters, using all mentions in them as input.

Specifically, the representation of a candidate mention $m_i$ is formed by concatenating the hidden states of its start and end tokens, followed by a fully connected (FC) layer, i.e., $\mathbf{m}_i=\text{FC}(\mathbf{h}_{i,s}\oplus\mathbf{h}_{i,e})\in \mathbb{R}^{d_h}$, where $\oplus$ denotes concatenation operator.
For historical clusters $\{c_j\}_{j=1}^w$ (where $f\leq g<i$), the representations of all mentions in each cluster are combined with $\mathbf{m}_i$ as follows:
\begin{align}
\mathbf{M}_j &= [\mathbf{m}_i,\mathbf{m}_{j,1}^f,\cdots,\mathbf{m}_{j,|c_j|}^g],\notag \\
    \mathbf{M}&=[\mathbf{M}_1; \cdots; \mathbf{M}_w],
\end{align}
where $\mathbf{M}\in \mathbb{R}^{w\times(\max_{j}(|c_j|)+1)\times d_h}$. Notably, clusters with fewer than $\max_j(|c_j|)$ mentions are padded with the [PAD] token's embedding. A Transformer (TF) encoder then processes $\mathbf{M}$, and the output of the [CLS] token is used:
\begin{align}
\mathbf{T}_{i}^{[CLS]}=\text{TF}(\mathbf{M}) \in \mathbb{R}^{w\times d_h}.
\label{eq:tf_cls}
\end{align}
Finally, an MLP classifier followed by a sigmoid function $\sigma$ computes the clustering probabilities:
\begin{align}
\mathbf{p}_c(m_i)=\sigma(\text{MLP}(\mathbf{T}_{i}^{[CLS]})) \in 
\mathbb{R}^{w\times 1}.
\end{align}
The mention $m_i$ is assigned to the cluster with the highest probability if $\max_j \mathbf{p}^j_c(m_i) > 0.5$; otherwise, a new cluster is created. The process is initialized with $c_1 = [m_1]$.

\section{Methodology} 
\subsection{Analytical Motivation}  
Although Eq.~(\ref{eq:tf_cls}) can model the global information of historical coreference clusters, the dimension of $\mathbf{M} \in \mathbb{R}^{w \times (\max_j(|c_j|)+1) \times d_h}$ expands rapidly as $w$ and $\max_j(|c_j|)$ increase. 
This leads to substantial growth in memory usage during training, severely limiting the model's practicality for long texts.
To illustrate this issue, Fig.~\ref{sta_onto_lit_fig:}(a) shows the distributions of $w$ and $\max_j(|c_j|)$ in the LitBank training corpus. 
It shows that many samples have large $w$ or $\max_j(|c_j|)$ values, directly inflating the dimension of $\mathbf{M}$.
Under computationally constrained settings, even using a lightweight Transformer model entails a non-negligible overhead, as shown in Fig.~\ref{sta_onto_lit_fig:}(b). See Section~\ref{sec:experiment} for more details. 

To address this, the proposed MEIC-DT approach employs a dual-threshold mechanism to control the input size of the lightweight Transformer by simultaneously limiting the number of historical coreference clusters and the number of mentions per cluster. 
Thereafter, we detail the design of MEIC-DT, including \textit{Statistics-Aware Eviction Strategy} and \textit{Internal Regularization Policy}.

\subsection{Statistics-Aware Eviction Strategy}
\label{sec:saes}
To retain crucial coreference clusters in a fixed-size cache, we propose a Statistics-Aware Eviction Strategy (\textbf{SAES}). 
SAES leverages the distinct information available in each phase: global statistics during training and real-time state during inference. 
This allows us to design differentiated, phase-specific eviction criteria.
Our strategy not only bounds computational cost by limiting the cache to at most $\tau_1$ clusters in the cache, where $\tau_1 > 0$ is a predefined integer threshold, but also improves the cache hit rate of target mentions.

Specifically, \textbf{\textit{during training}}, where ground-truth coreference cluster annotations are available, we construct a ranking score for each tracked cluster $c_j$ ($j \in [1, \tau_1]$) based on its number of remaining mentions $rm_j$ and the number of steps since its last access $lru_j$:
\begin{align}
    SAES_{\text{train},j} = rm_j \cdot \left(1+\frac{1}{lru_j + \delta}\right),
\end{align}
where $\delta$ is a very small positive scalar (default \(1 \times 10^{-5}\)). 
When the model's clustering confidence for the current mention $m_i$ is insufficient (i.e., $\max_j \mathbf{p}^j_c(m_i) \leq 0.5$) and the number of tracked clusters reaches $\tau_1$, the cluster with the smallest $SAES_{\text{train},j}$ is evicted. 
Otherwise, a new cluster $c_{w+1}=[m_i]$ is created.
This design prioritizes clusters with more remaining mentions (i.e., dominated by $rm_j$), which helps maintain long-distance coreference links.  
The second term acts as a tie-breaker: when $rm_j$ is tied, a larger $lru_j$ results in a lower score, enforcing LRU eviction.
When $lru_j = 0$ (i.e., the cluster was just accessed), the second term is maximized, giving that cluster the highest protection from eviction.

\textbf{\textit{During inference}}, the true $rm_j$ value is unavailable. Relying solely on LRU may mistakenly remove important clusters. Therefore, we introduce a composite score that combines the number of existing mentions $em_j$ in tracked cluster $c_j$, the total number of steps since its creation $age_j$, and $lru_j$:
\begin{align}
    SAES_{\text{inf},j} = \frac{em_j}{age_j} \cdot \left(1+ \frac{1}{lru_j +\delta}\right). 
\end{align}
Here, the ratio $\frac{em_j}{age_j}$ reflects the tracked cluster's activity level---a higher value suggests greater coreference potential. Similarly, clusters are sorted by $SAES_{\text{inf},j}$ in ascending order. When clustering confidence is low and the number of tracked clusters reaches $\tau_1$, the cluster with the smallest score is evicted.
Moreover, the second term ensures that among clusters with identical mention activity level, the one with the largest $lru_j$ (least recently used) is evicted.

\underline{\textit{\textbf{Comment 1.}}} Unlike prior methods that rely solely on LRU~\cite{xia-etal-2020-incremental, toshniwal-etal-2020-learning} or complex dual-cache mechanisms~\cite{guo-etal-2023-dual}, SAES employs a unified yet statistics-aware scoring function. It utilizes richer information (global mention counts in training, activity rates in inference) while maintaining a simple, single-cache architecture, thus simplifying management.


\subsection{Internal Regularization Policy}
\label{sec:irp}
To select representative mentions under constrained memory, we introduce an Internal Regularization Policy (\textbf{IRP}) that operates on each tracked coreference cluster. 
This policy limits the number of mentions stored per cluster to a predefined positive integer threshold $\tau_2$. Clusters with at most $\tau_2$ mentions are left unchanged; larger clusters are condensed to their $\tau_2$ most representative mentions.
The key challenge is to select a subset of mentions that best preserves the cluster’s semantic information.
To achieve this, we craft a two-stage algorithm: (1) a grouping strategy based on mention textual relations, and (2) a group-based sampling for selecting representative mentions.

Specifically, \textbf{\textit{in the first stage}}, we use a Union-Find algorithm to efficiently cluster semantically similar mentions based on textual equivalence/inclusion relationships and a pronoun lexicon (see Appendix~\ref{app:full_pronouns_list}), forming groups that serve as the sampling units.
\textbf{\textit{In the second stage}}, we adopt a group-based sampling strategy. It deterministically retains the first and last mentions of the original cluster to preserve contextual boundaries. For the remaining $\tau_2 - 2$ slots, quotas are dynamically allocated across groups: if the number of groups exceeds available slots, the largest groups are prioritized; otherwise, at least one mention is selected from each group. This strategy maximizes the coverage of diverse semantic units within the condensed cluster. Pseudocode is provided in Appendix~\ref{app:code_IRP}.

\underline{\textit{\textbf{Comment 2.}}} 
The two-stage design of IRP effectively selects a compact yet representative set of mentions from the original cluster. Compared to the existing Transformer-based incremental model~\cite{martinelli-etal-2024-maverick}, \textbf{\textit{IRP is, to our knowledge, a novel design that effectively maintains CR performance under strict memory constraints through its structured information preservation policy.}}

\underline{\textit{\textbf{Comment 3.}}} Together, IRP and SAES ensure that the representation $\mathbf{M}$ is bounded in size by $\mathbb{R}^{\tau_1 \times (\tau_2+1) \times d_h}$, guaranteeing that the incremental clustering process runs efficiently within the predefined memory budget.

\section{Experiments}
\label{sec:experiment}

\subsection{Experimental Settings}

\textbf{Datasets.} We train and evaluate the performance of our proposed approach on two widely used CR datasets: OntoNotes~\cite{pradhan-etal-2012-conll} and LitBank~\cite{bamman-etal-2020-annotated}. 
To assess the generalization capability of approaches, models trained on OntoNotes are also tested on the out-of-domain dataset WikiCoref~\cite{ghaddar-langlais-2016-wikicoref}. 
Statistics are detailed in Table~\ref{tab:Dataset_statistics}.
\begin{table}[htbp]
  \centering
  \resizebox{1.0\columnwidth}{!}{
    \begin{tabular}{l|ccccc}
    \toprule
    Datasets & \#Train & \#Val & \#Test & Avg. W & Avg. M  \\
    \hline
    OntoNotes & 2802  & 343   & 348   & 467   & 56      \\
    LitBank & 80    & 10    & 10    & 2105  & 291     \\
    WikiCoref & -     & -     & 30    & 1996  & 230    \\ 
    \bottomrule
    \end{tabular}%
    }
    \caption{Dataset statistics: number of documents in each dataset split (\#Train/Val/Test), average number of words (i.e., Avg. W) and mentions (i.e., Avg. M) per document.}
  \label{tab:Dataset_statistics}%
  \vspace*{-2ex}
\end{table}%

\textbf{Baselines.} As outlined in Related Works (see Appendix~\ref{Related_works:app}), while numerous CR approaches exist, our work specifically focuses on memory efficiency for incremental clustering. 
Accordingly, we select the most relevant baselines in the main paper: LRU~\cite{xia-etal-2020-incremental, toshniwal-etal-2020-learning} and Dual-cache~\cite{guo-etal-2023-dual}. For more comparison results, please see Appendix~\ref{app:add_exp_res}. 
To gauge the performance of MEIC-DT (which integrates SAES and IRP), we follow~\citet{luo2025imcoref} in adopting Imcoref as Mention Extractor and DeBERTa$_{\text{large}}$~\cite{he2021deberta} as Text Encoder~(see Fig.~\ref{Framework_fig:}). Furthermore, we conduct coreference clustering experiments adopting two distinct classifiers---a linear classifier~\cite{xia-etal-2020-incremental, toshniwal-etal-2020-learning,guo-etal-2023-dual} and a lightweight Transformer classifier~\cite{martinelli-etal-2024-maverick}---to thoroughly evaluate the effectiveness of MEIC-DT.

\textbf{Metrics.}
In our experiments, we evaluate performance using the F1 scores of the MUC~\cite{vilain1995model}, B$^3$~\cite{bagga1998algorithms}, and CEAF$_{\phi_4}$~\cite{luo2005coreference} metrics. The overall CoNLL-F1 score (abbreviated as Avg.F1) is the average of these three F1 scores. To ensure reliability, we report the average for each experiment over $3$ different random seeds.

\textbf{Configurations.}
Unless noted otherwise, we set $\tau_1=50$ and $\tau_2=30$ for the Transformer classifier, and $\tau_1=50$ for the linear classifier. For the linear classifier, we follow prior works~\citet{xia-etal-2020-incremental,toshniwal-etal-2020-learning,guo-etal-2023-dual} in representing each historical cluster with a single vector, which eliminates the need for additional parameters beyond $\tau_1$. 
We set $\delta=1\times 10^{-5}$ and employ the Adafactor optimizer~\cite{shazeer2018adafactor} for model training, with learning rates configured as 2e-5 for DeBERTa$_{\text{large}}$ and 3e-4 for the remaining model layers.
All experiments are implemented using the PyTorch-Lightning framework. Each run is executed on a single RTX 4090 GPU with 24GB of VRAM.  
Due to space limitations, the complete experimental settings are provided in Appendix~\ref{app:com_exp_set}.

\begin{table*}[ht]
  \centering
  \resizebox{2.0\columnwidth}{!}{
    \begin{tabular}{cccccc|cccc|cccc}
    \toprule
    \multirow{2}[1]{*}{\textbf{Methods}} & \multirow{2}[1]{*}{$\tau_1$/$\tau_2$} & \multicolumn{4}{c|}{\textbf{Litbank}}  & \multicolumn{4}{c|}{\textbf{OntoNotes}} & \multicolumn{4}{c}{\textbf{WikiCoref}} \\
          &       & \textbf{MUC}   & \textbf{B}$^3$   & \textbf{CEAF$_{\phi_4}$} & \textbf{\textcolor[rgb]{0.8,0,0}{Avg.F1}} & \textbf{MUC}   & \textbf{B}$^3$   & \textbf{CEAF$_{\phi_4}$} & \textbf{\textcolor[rgb]{0.8,0,0}{Avg.F1}} & \textbf{MUC}   & \textbf{B}$^3$   & \textbf{CEAF$_{\phi_4}$} & \textbf{\textcolor[rgb]{0.8,0,0}{Avg.F1}} \\
    \rowcolor{mypink}\multicolumn{14}{c}{\textbf{Linear Classifier}} \\
    Unbound & \multicolumn{1}{l}{-/-} & 84.52$_{\pm0.77}$ & 65.67$_{\pm0.87}$ & 76.59$_{\pm0.37}$ & 75.59$_{\pm0.86}$ & 89.71$_{\pm0.28}$ & 77.31$_{\pm0.68}$ & 69.88$_{\pm0.72}$ & 78.97$_{\pm0.58}$ & 72.19$_{\pm0.38}$ & 53.34$_{\pm0.20}$ & 64.28$_{\pm0.59}$ & 63.27$_{\pm0.12}$ \\
    \hline
    \multirow{5}[2]{*}{\textbf{LRU}} & \multicolumn{1}{l}{10/-} & 83.77$_{\pm0.37}$ & 66.76$_{\pm0.40}$ & 74.94$_{\pm0.27}$ & 75.16$_{\pm0.27}$ & 87.94$_{\pm0.17}$ & 75.15$_{\pm0.43}$ & 66.15$_{\pm0.74}$ & 76.41$_{\pm0.91}$ & 71.43$_{\pm0.09}$ & 54.43$_{\pm0.19}$ & 62.62$_{\pm0.73}$ & 62.83$_{\pm0.99}$ \\
          & \multicolumn{1}{l}{30/-} & 83.91$_{\pm0.19}$ & 72.38$_{\pm0.83}$ & 71.37$_{\pm0.14}$ & 75.89$_{\pm0.32}$ & 88.49$_{\pm0.38}$ & 76.05$_{\pm0.92}$ & 68.05$_{\pm1.84}$ & 77.53$_{\pm1.02}$ & 73.01$_{\pm0.63}$ & 53.80$_{\pm0.42}$ & 64.08$_{\pm0.98}$ & 63.63$_{\pm0.80}$ \\
          & \multicolumn{1}{l}{50/-} & 84.96$_{\pm0.73}$ & 66.27$_{\pm0.91}$ & 77.55$_{\pm0.36}$ & \textbf{76.26$_{\pm0.17}$} & 87.68$_{\pm0.34}$ & 75.48$_{\pm0.75}$ & 71.28$_{\pm0.44}$ & 78.15$_{\pm0.47}$ & 72.41$_{\pm0.34}$ & 52.80$_{\pm0.43}$ & 65.03$_{\pm0.50}$ & 63.42$_{\pm0.63}$ \\
          & \multicolumn{1}{l}{100/-} & 84.76$_{\pm1.38}$ & 65.14$_{\pm0.82}$ & 77.35$_{\pm1.20}$ & 75.75$_{\pm0.97}$ & 88.76$_{\pm0.31}$ & 75.85$_{\pm0.31}$ & 71.14$_{\pm0.17}$ & 78.58$_{\pm0.13}$ & 72.35$_{\pm0.10}$ & 53.95$_{\pm0.01}$ & 64.23$_{\pm0.92}$ & 63.51$_{\pm0.83}$ \\
          & \multicolumn{1}{l}{200/-} & 84.52$_{\pm0.77}$ & 65.67$_{\pm0.87}$ & 76.59$_{\pm1.37}$ & 75.59$_{\pm0.86}$ & 8.68$_{\pm0.37}$ & 75.78$_{\pm0.75}$ & 71.48$_{\pm0.55}$ & \textbf{78.65$_{\pm0.36}$} & 72.65$_{\pm0.82}$ & 53.94$_{\pm0.67}$ & 65.21$_{\pm0.75}$ & \textbf{63.93$_{\pm0.43}$} \\
    \hline
    \multirow{5}[2]{*}{\textbf{Dual-cache}} & \multicolumn{1}{l}{10/-} & 84.06$_{\pm1.40}$ & 66.63$_{\pm0.72}$ & 73.71$_{\pm0.46}$ & 74.80$_{\pm0.19}$ & 87.23$_{\pm0.81}$ & 75.18$_{\pm1.00}$ & 70.50$_{\pm0.46}$ & 77.64$_{\pm0.75}$ & 71.68$_{\pm0.50}$ & 54.26$_{\pm0.85}$ & 61.34$_{\pm0.99}$ & 62.43$_{\pm0.18}$ \\
          & \multicolumn{1}{l}{30/-} & 85.26$_{\pm0.34}$ & 66.63$_{\pm0.72}$ & 76.73$_{\pm0.81}$ & 76.21$_{\pm0.08}$ & 87.19$_{\pm0.69}$ & 75.26$_{\pm0.94}$ & 71.42$_{\pm0.60}$ & 77.96$_{\pm0.65}$ & 72.86$_{\pm0.13}$ & 54.25$_{\pm0.04}$ & 64.37$_{\pm0.87}$ & 63.83$_{\pm0.05}$ \\
          & \multicolumn{1}{l}{50/-} & 85.42$_{\pm0.55}$ & 66.46$_{\pm0.98}$ & 78.20$_{\pm0.27}$ & \textbf{76.69$_{\pm0.52}$} & 88.54$_{\pm0.42}$ & 76.39$_{\pm0.38}$ & 71.63$_{\pm0.75}$ & \textbf{78.85$_{\pm0.56}$} & 73.04$_{\pm0.68}$ & 54.1$_{\pm0.26}$ & 65.82$_{\pm0.76}$ & \textbf{64.32$_{\pm0.90}$} \\
          & \multicolumn{1}{l}{100/-} & 84.22$_{\pm0.43}$ & 65.36$_{\pm0.98}$ & 77.01$_{\pm0.27}$ & 75.53$_{\pm0.52}$ & 88.21$_{\pm0.32}$ & 75.41$_{\pm0.31}$ & 71.31$_{\pm0.42}$ & 78.31$_{\pm0.31}$ & 71.82$_{\pm0.72}$ & 53.01$_{\pm0.39}$ & 64.65$_{\pm0.47}$ & 63.16$_{\pm0.16}$ \\
          & \multicolumn{1}{l}{200/-} & 84.52$_{\pm0.77}$ & 65.67$_{\pm0.87}$ & 76.59$_{\pm0.37}$ & 75.59$_{\pm0.86}$ & 88.27$_{\pm0.44}$ & 75.86$_{\pm0.39}$ & 71.81$_{\pm0.76}$ & 78.65$_{\pm0.53}$ & 73.45$_{\pm0.76}$ & 52.98$_{\pm0.89}$ & 64.79$_{\pm0.65}$ & 63.74$_{\pm0.75}$ \\
    \hline
    \multirow{10}[11]{*}{\textbf{MEIC-DT}} & \multicolumn{1}{l}{10/-} & 84.71$_{\pm0.10}$ & 65.84$_{\pm0.15}$ & 76.83$_{\pm0.19}$ & 75.79$_{\pm0.75}$ & 87.17$_{\pm0.39}$ & 75.49$_{\pm0.73}$ & 70.76$_{\pm0.21}$ & 77.81$_{\pm0.63}$ & 71.31$_{\pm0.19}$ & 52.69$_{\pm0.75}$ & 64.07$_{\pm0.73}$ & 62.69$_{\pm0.30}$ \\
          & \cellcolor{blue!5}$\Delta$ & \cellcolor{blue!5}\textcolor[rgb]{0.7,0,0}{+0.79}  & \cellcolor{blue!5}\textcolor[rgb]{0,0.7,0}{-0.85}  & \cellcolor{blue!5}\textcolor[rgb]{0.7,0,0}{2.51}  & \cellcolor{blue!5}\textcolor[rgb]{0.7,0,0}{+0.81}  & \cellcolor{blue!5}\textcolor[rgb]{0,0.7,0}{-0.41}  & \cellcolor{blue!5}\textcolor[rgb]{0.7,0,0}{+0.32}  & \cellcolor{blue!5}\textcolor[rgb]{0.7,0,0}{+2.44}  & \cellcolor{blue!5}\textcolor[rgb]{0.7,0,0}{+0.79}  & \cellcolor{blue!5}\textcolor[rgb]{0,0.7,0}{-0.25}  & \cellcolor{blue!5}\textcolor[rgb]{0,0.7,0}{-1.66}  & \cellcolor{blue!5}\textcolor[rgb]{0.7,0,0}{+2.09}  & \cellcolor{blue!5}\textcolor[rgb]{0.7,0,0}{+0.06}  \\
\cdashline{2-14}          & \multicolumn{1}{l}{30/-} & 84.98$_{\pm0.54}$ & 67.49$_{\pm0.59}$ & 77.20$_{\pm1.00}$ & 76.56$_{\pm0.96}$ & 87.13$_{\pm0.48}$ & 76.06$_{\pm0.95}$ & 71.17$_{\pm0.66}$ &78.12$_{\pm0.64}$ & 72.44$_{\pm0.46}$ & 54.97$_{\pm0.42}$ & 64.65$_{\pm0.49}$ & 64.02$_{\pm0.67}$ \\
          & \cellcolor{blue!5}$\Delta$ & \cellcolor{blue!5}\textcolor[rgb]{0,0.7,0}{-0.33}  & \cellcolor{blue!5}\textcolor[rgb]{0.7,0,0}{+1.12}  & \cellcolor{blue!5}\textcolor[rgb]{0.7,0,0}{+0.63}  & \cellcolor{blue!5}\textcolor[rgb]{0.7,0,0}{+0.48}  & \cellcolor{blue!5}\textcolor[rgb]{0,0.7,0}{-0.71}  & \cellcolor{blue!5}\textcolor[rgb]{0.7,0,0}{+0.41}  & \cellcolor{blue!5}\textcolor[rgb]{0.7,0,0}{+1.44}  & \cellcolor{blue!5}\textcolor[rgb]{0.7,0,0}{+0.38}  & \cellcolor{blue!5}\textcolor[rgb]{0,0.7,0}{-0.50}  & \cellcolor{blue!5}\textcolor[rgb]{0.7,0,0}{+0.95}  & \cellcolor{blue!5}\textcolor[rgb]{0.7,0,0}{+0.43}  & \cellcolor{blue!5}\textcolor[rgb]{0.7,0,0}{+0.29}  \\
\cdashline{2-14}          & \multicolumn{1}{l}{50/-} & 85.41$_{\pm0.44}$ & 67.57$_{\pm0.14}$ & 78.15$_{\pm0.65}$ & \underline{\textbf{77.04$_{\pm0.28}$}} & 87.53$_{\pm0.37}$ & 77.58$_{\pm0.42}$ & 72.85$_{\pm0.88}$ & \underline{\textbf{79.32$_{\pm0.67}$}} & 72.84$_{\pm0.43}$ & 55.06$_{\pm0.42}$ & 65.64$_{\pm0.47}$ & \underline{\textbf{64.51$_{\pm0.49}$}} \\
          & \cellcolor{blue!5}$\Delta$ & \cellcolor{blue!5}\textcolor[rgb]{0.7,0,0}{+0.22}  & \cellcolor{blue!5}\textcolor[rgb]{0.7,0,0}{+1.21}  & \cellcolor{blue!5}\textcolor[rgb]{0.7,0,0}{+0.28}  & \cellcolor{blue!5}\textcolor[rgb]{0.7,0,0}{+0.57}  & \cellcolor{blue!5}\textcolor[rgb]{0,0.7,0}{-0.58}  & \cellcolor{blue!5}\textcolor[rgb]{0.7,0,0}{+1.65}  & \cellcolor{blue!5}\textcolor[rgb]{0.7,0,0}{+1.40}  & \cellcolor{blue!5}\textcolor[rgb]{0.7,0,0}{+0.82}  & \cellcolor{blue!5}\textcolor[rgb]{0.7,0,0}{+0.12}  & \cellcolor{blue!5}\textcolor[rgb]{0.7,0,0}{+1.61}  & \cellcolor{blue!5}\textcolor[rgb]{0.7,0,0}{+0.22}  & \cellcolor{blue!5}\textcolor[rgb]{0.7,0,0}{+0.64}  \\
\cdashline{2-14}          & \multicolumn{1}{l}{100/-} & 85.56$_{\pm0.05}$ & 67.46$_{\pm0.11}$ & 78.02$_{\pm0.42}$ & 77.01$_{\pm0.46}$ & 87.95$_{\pm0.42}$ & 76.17$_{\pm0.41}$ & 72.96$_{\pm0.85}$ & 79.03$_{\pm0.59}$ & 73.02$_{\pm0.19}$ & 54.94$_{\pm0.27}$ & 65.50$_{\pm0.22}$ & 64.49$_{\pm0.41}$ \\
          & \cellcolor{blue!5}$\Delta$ & \cellcolor{blue!5}\textcolor[rgb]{0.7,0,0}{+1.07}  & \cellcolor{blue!5}\textcolor[rgb]{0.7,0,0}{+2.21}  & \cellcolor{blue!5}\textcolor[rgb]{0.7,0,0}{+0.84}  & \cellcolor{blue!5}\textcolor[rgb]{0.7,0,0}{+1.37}  & \cellcolor{blue!5}\textcolor[rgb]{0,0.7,0}{-0.53}  & \cellcolor{blue!5}\textcolor[rgb]{0.7,0,0}{+0.54}  & \cellcolor{blue!5}\textcolor[rgb]{0.7,0,0}{+1.73}  & \cellcolor{blue!5}\textcolor[rgb]{0.7,0,0}{+0.59}  & \cellcolor{blue!5}\textcolor[rgb]{0.7,0,0}{+0.94}  & \cellcolor{blue!5}\textcolor[rgb]{0.7,0,0}{+1.46}  & \cellcolor{blue!5}\textcolor[rgb]{0.7,0,0}{+1.06}  & \cellcolor{blue!5}\textcolor[rgb]{0.7,0,0}{+1.16}  \\
\cdashline{2-14}          & \multicolumn{1}{l}{200/-} & 84.52$_{\pm0.77}$ & 65.67$_{\pm0.87}$ & 76.59$_{\pm0.37}$ & 75.59$_{\pm0.86}$ & 87.38$_{\pm0.27}$ & 76.37$_{\pm0.54}$ & 72.41$_{\pm0.69}$ & 78.72$_{\pm0.43}$ & 73.53$_{\pm0.72}$ & 54.65$_{\pm0.62}$ & 65.32$_{\pm0.71}$ & 64.50$_{\pm0.68}$ \\
          & \cellcolor{blue!5}$\Delta$ & \cellcolor{blue!5}\textcolor[rgb]{0.7,0,0}{+0.00}  & \cellcolor{blue!5}\textcolor[rgb]{0.7,0,0}{+0.00}  & \cellcolor{blue!5}\textcolor[rgb]{0.7,0,0}{+0.00}  & \cellcolor{blue!5}\textcolor[rgb]{0.7,0,0}{+0.00}  & \cellcolor{blue!5}\textcolor[rgb]{0,0.7,0}{-1.10}  & \cellcolor{blue!5}\textcolor[rgb]{0.7,0,0}{+0.55}  & \cellcolor{blue!5}\textcolor[rgb]{0.7,0,0}{+0.76}  & \cellcolor{blue!5}\textcolor[rgb]{0.7,0,0}{+0.07}  & \cellcolor{blue!5}\textcolor[rgb]{0.7,0,0}{+0.48}  & \cellcolor{blue!5}\textcolor[rgb]{0.7,0,0}{+1.19}  & \cellcolor{blue!5}\textcolor[rgb]{0.7,0,0}{+0.32}  & \cellcolor{blue!5}\textcolor[rgb]{0.7,0,0}{+0.66}  \\
    \rowcolor{mypink}\multicolumn{14}{c}{\textbf{Transformer Classifier}} \\
    \multirow{5}[1]{*}{\textbf{LRU+IRP}} & \multicolumn{1}{l}{10/10} & 84.54$_{\pm0.39}$ & 70.29$_{\pm0.14}$ & 72.25$_{\pm0.64}$ & 75.69$_{\pm0.63}$ & 87.82$_{\pm0.43}$ & 80.50$_{\pm0.37}$ & 73.09$_{\pm0.12}$ & 80.47$_{\pm0.62}$ & 77.63$_{\pm0.68}$ & 63.38$_{\pm0.61}$ & 65.34$_{\pm0.38}$ & 68.78$_{\pm0.89}$ \\
          & \multicolumn{1}{l}{10/20} & 85.36$_{\pm0.76}$ & 66.11$_{\pm0.94}$ & 76.40$_{\pm1.24}$ & 75.96$_{\pm0.97}$ & 87.75$_{\pm0.01}$ & 80.80$_{\pm0.26}$ & 74.82$_{\pm0.69}$ & 81.12$_{\pm0.33}$ & 76.97$_{\pm0.10}$ & 65.44$_{\pm0.26}$ & 64.47$_{\pm0.39}$ & 68.96$_{\pm0.08}$ \\
          & \multicolumn{1}{l}{10/30} & 84.70$_{\pm0.24}$ & 73.69$_{\pm0.42}$ & 71.90$_{\pm0.92}$ & 76.76$_{\pm0.49}$ & 88.25$_{\pm0.17}$ & 81.10$_{\pm0.92}$ & 78.74$_{\pm0.21}$ & 82.70$_{\pm0.64}$ & 77.79$_{\pm0.79}$ & 66.74$_{\pm0.41}$ & 64.96$_{\pm0.79}$ & 69.83$_{\pm0.66}$ \\
          & \multicolumn{1}{l}{30/30} & 86.39$_{\pm0.21}$ & 74.44$_{\pm0.63}$ & 75.22$_{\pm0.80}$ & 78.68$_{\pm0.59}$ & 89.15$_{\pm0.15}$ & 82.02$_{\pm0.18}$ & 79.04$_{\pm0.76}$ & 83.40$_{\pm0.22}$ & 78.89$_{\pm0.31}$ & 64.56$_{\pm0.48}$ & 67.93$_{\pm0.36}$ & 70.46$_{\pm0.73}$ \\
          & \multicolumn{1}{l}{50/30} & 86.68$_{\pm0.71}$ & 74.96$_{\pm0.13}$ & 75.95$_{\pm0.64}$ & \textbf{79.19$_{\pm0.21}$} & 89.24$_{\pm0.05}$ & 82.57$_{\pm0.49}$ & 79.43$_{\pm0.83}$ & \textbf{83.75$_{\pm0.36}$} & 78.92$_{\pm0.35}$ & 65.65$_{\pm0.68}$ & 67.93$_{\pm0.49}$ & \textbf{70.83$_{\pm0.65}$} \\
    \hline
    \multirow{5}[2]{*}{\textbf{Dual-cache+IRP}} & \multicolumn{1}{l}{10/10} & 85.40$_{\pm0.18}$ & 71.81$_{\pm0.21}$ & 73.52$_{\pm0.64}$ & 76.91$_{\pm0.89}$ & 88.41$_{\pm0.34}$ & 81.11$_{\pm0.78}$ & 73.25$_{\pm0.69}$ & 80.92$_{\pm0.59}$ & 78.46$_{\pm0.29}$ & 64.88$_{\pm0.46}$ & 66.57$_{\pm0.42}$ & 69.97$_{\pm0.35}$ \\
          & \multicolumn{1}{l}{10/20} & 85.37$_{\pm0.54}$ & 72.59$_{\pm0.28}$ & 74.43$_{\pm0.93}$ & 77.46$_{\pm0.12}$ & 88.84$_{\pm0.48}$ & 81.74$_{\pm0.37}$ & 75.96$_{\pm0.65}$ & 82.18$_{\pm0.44}$ & 78.37$_{\pm0.15}$ & 65.61$_{\pm0.24}$ & 67.45$_{\pm0.43}$ & 70.48$_{\pm0.18}$ \\
          & \multicolumn{1}{l}{10/30} & 86.20$_{\pm0.94}$ & 74.37$_{\pm0.84}$ & 73.90$_{\pm0.43}$ & 78.16$_{\pm0.38}$ & 89.19$_{\pm0.41}$ & 83.52$_{\pm0.38}$ & 76.56$_{\pm0.63}$ & 83.09$_{\pm0.37}$ & 79.21$_{\pm0.31}$ & 64.89$_{\pm0.09}$ & 68.21$_{\pm0.63}$ & 70.74$_{\pm0.11}$ \\
          & \multicolumn{1}{l}{30/30} & 86.89$_{\pm0.10}$ & 74.79$_{\pm0.11}$ & 76.03$_{\pm0.59}$ & 79.24$_{\pm0.15}$ & 90.18$_{\pm0.10}$ & 84.27$_{\pm0.44}$ & 77.35$_{\pm0.56}$ & 83.93$_{\pm0.27}$ & 79.17$_{\pm0.09}$ & 66.67$_{\pm0.46}$ & 67.83$_{\pm0.61}$ & 71.22$_{\pm0.32}$ \\
          & \multicolumn{1}{l}{50/30} & 87.40$_{\pm0.39}$ & 77.04$_{\pm0.16}$ & 77.40$_{\pm0.73}$ & \textbf{80.61$_{\pm0.71}$} & 90.29$_{\pm0.46}$ & 83.30$_{\pm0.53}$ & 80.72$_{\pm0.97}$ & \textbf{84.77$_{\pm0.64}$} & 79.73$_{\pm0.24}$ & 64.30$_{\pm0.20}$ & 69.83$_{\pm0.41}$ & \textbf{71.29$_{\pm0.27}$} \\
    \hline
    \multirow{10}[10]{*}{\textbf{MEIC-DT}} & \multicolumn{1}{l}{10/10} & 86.09$_{\pm0.94}$ & 72.66$_{\pm0.18}$ & 75.05$_{\pm0.20}$ & 77.93$_{\pm0.76}$ & 88.59$_{\pm0.99}$ & 81.58$_{\pm0.14}$ & 74.07$_{\pm0.89}$ & 81.42$_{\pm0.89}$ & 79.13$_{\pm0.24}$ & 65.69$_{\pm0.23}$ & 68.08$_{\pm0.41}$ & 70.97$_{\pm0.26}$ \\
          & \cellcolor{blue!5}$\Delta$ & \cellcolor{blue!5}\textcolor[rgb]{0.7,0,0}{+1.12}  & \cellcolor{blue!5}\textcolor[rgb]{0.7,0,0}{+1.61}  & \cellcolor{blue!5}\textcolor[rgb]{0.7,0,0}{+2.17}  & \cellcolor{blue!5}\textcolor[rgb]{0.7,0,0}{+1.63}  & \cellcolor{blue!5}\textcolor[rgb]{0.7,0,0}{+0.48}  & \cellcolor{blue!5}\textcolor[rgb]{0.7,0,0}{+0.77}  & \cellcolor{blue!5}\textcolor[rgb]{0.7,0,0}{+0.90}  & \cellcolor{blue!5}\textcolor[rgb]{0.7,0,0}{+0.73}  & \cellcolor{blue!5}\textcolor[rgb]{0.7,0,0}{+1.09}  & \cellcolor{blue!5}\textcolor[rgb]{0.7,0,0}{+1.56}  & \cellcolor{blue!5}\textcolor[rgb]{0.7,0,0}{+2.13}  & \cellcolor{blue!5}\textcolor[rgb]{0.7,0,0}{+1.60}  \\
\cdashline{2-14}          & \multicolumn{1}{l}{10/20} & 86.46$_{\pm0.94}$ & 72.63$_{\pm0.16}$ & 75.52$_{\pm0.40}$ & 78.20$_{\pm0.16}$ & 89.23$_{\pm0.58}$ & 81.75$_{\pm0.19}$ & 79.48$_{\pm1.00}$ & 83.49$_{\pm0.46}$ & 79.33$_{\pm0.24}$ & 65.50$_{\pm0.53}$ & 68.40$_{\pm0.39}$ & 71.08$_{\pm0.26}$ \\
          & \cellcolor{blue!5}$\Delta$ & \cellcolor{blue!5}\textcolor[rgb]{0.7,0,0}{+1.82}  & \cellcolor{blue!5}\textcolor[rgb]{0.7,0,0}{+0.15}  & \cellcolor{blue!5}\textcolor[rgb]{0.7,0,0}{+2.62}  & \cellcolor{blue!5}\textcolor[rgb]{0.7,0,0}{+1.52}  & \cellcolor{blue!5}\textcolor[rgb]{0.7,0,0}{+0.94}  & \cellcolor{blue!5}\textcolor[rgb]{0.7,0,0}{+0.48}  & \cellcolor{blue!5}\textcolor[rgb]{0.7,0,0}{+4.09}  & \cellcolor{blue!5}\textcolor[rgb]{0.7,0,0}{+1.84}  & \cellcolor{blue!5}\textcolor[rgb]{0.7,0,0}{+1.66}  & \cellcolor{blue!5}\textcolor[rgb]{0,0.7,0}{-0.02}  & \cellcolor{blue!5}\textcolor[rgb]{0.7,0,0}{+2.44}  & \cellcolor{blue!5}\textcolor[rgb]{0.7,0,0}{+1.36}  \\
\cdashline{2-14}          & \multicolumn{1}{l}{10/30} & 86.55$_{\pm0.33}$ & 73.92$_{\pm0.91}$ & 75.37$_{\pm0.65}$ & 78.61$_{\pm0.92}$ & 89.40$_{\pm0.89}$ & 82.40$_{\pm0.40}$ & 79.58$_{\pm0.39}$ & 83.80$_{\pm0.56}$ & 79.41$_{\pm0.38}$ & 66.73$_{\pm0.45}$ & 68.21$_{\pm0.32}$ & 71.45$_{\pm0.31}$ \\
          & \cellcolor{blue!5}$\Delta$ & \cellcolor{blue!5}\textcolor[rgb]{0.7,0,0}{+1.10}  & \cellcolor{blue!5}\textcolor[rgb]{0,0.7,0}{-0.11}  & \cellcolor{blue!5}\textcolor[rgb]{0.7,0,0}{+2.47}  & \cellcolor{blue!5}\textcolor[rgb]{0.7,0,0}{+1.15}  & \cellcolor{blue!5}\textcolor[rgb]{0.7,0,0}{+0.68}  & \cellcolor{blue!5}\textcolor[rgb]{0.7,0,0}{+0.09}  & \cellcolor{blue!5}\textcolor[rgb]{0.7,0,0}{+1.93}  & \cellcolor{blue!5}\textcolor[rgb]{0.7,0,0}{+0.90}  & \cellcolor{blue!5}\textcolor[rgb]{0.7,0,0}{+0.91}  & \cellcolor{blue!5}\textcolor[rgb]{0.7,0,0}{+0.92}  & \cellcolor{blue!5}\textcolor[rgb]{0.7,0,0}{+1.63}  & \cellcolor{blue!5}\textcolor[rgb]{0.7,0,0}{+1.17}  \\
\cdashline{2-14}          & \multicolumn{1}{l}{30/30} & 86.86$_{\pm0.18}$ & 74.77$_{\pm0.99}$ & 75.91$_{\pm0.79}$ & 79.18$_{\pm0.08}$ & 90.05$_{\pm0.46}$ & 84.49$_{\pm0.31}$ & 75.24$_{\pm0.16}$ & 84.26$_{\pm0.63}$ & 79.73$_{\pm0.18}$ & 67.69$_{\pm0.33}$ & 68.90$_{\pm0.60}$ & 72.11$_{\pm0.38}$ \\
          & \cellcolor{blue!5}$\Delta$ & \cellcolor{blue!5}\textcolor[rgb]{0.7,0,0}{+0.22}  & \cellcolor{blue!5}\textcolor[rgb]{0.7,0,0}{+0.16}  & \cellcolor{blue!5}\textcolor[rgb]{0.7,0,0}{+0.28}  & \cellcolor{blue!5}\textcolor[rgb]{0.7,0,0}{+0.22}  & \cellcolor{blue!5}\textcolor[rgb]{0.7,0,0}{+0.38}  & \cellcolor{blue!5}\textcolor[rgb]{0.7,0,0}{+1.35}  & \cellcolor{blue!5}\textcolor[rgb]{0.7,0,0}{+0.04}  & \cellcolor{blue!5}\textcolor[rgb]{0.7,0,0}{+0.59}  & \cellcolor{blue!5}\textcolor[rgb]{0.7,0,0}{+0.70}  & \cellcolor{blue!5}\textcolor[rgb]{0.7,0,0}{+2.08}  & \cellcolor{blue!5}\textcolor[rgb]{0.7,0,0}{+1.02}  & \cellcolor{blue!5}\textcolor[rgb]{0.7,0,0}{+1.27}  \\
\cdashline{2-14}          & \multicolumn{1}{l}{50/30} & 87.85$_{\pm0.28}$ & 78.53$_{\pm0.24}$ & 77.42$_{\pm0.03}$ & \underline{\textbf{81.27$_{\pm0.41}$}} & 90.53$_{\pm0.43}$ & 84.43$_{\pm0.42}$ & 80.19$_{\pm0.49}$ & \underline{\textbf{85.05$_{\pm0.67}$}} & 80.16$_{\pm0.25}$ & 68.38$_{\pm0.21}$ & 69.35$_{\pm0.43}$ & \underline{\textbf{72.63$_{\pm0.24}$}} \\
          & \cellcolor{blue!5}$\Delta$ & \cellcolor{blue!5}\textcolor[rgb]{0.7,0,0}{+0.31}  & \cellcolor{blue!5}\textcolor[rgb]{0.7,0,0}{+2.53}  & \cellcolor{blue!5}\textcolor[rgb]{0,0.7,0}{-0.26}  & \cellcolor{blue!5}\textcolor[rgb]{0.7,0,0}{+0.87}  & \cellcolor{blue!5}\textcolor[rgb]{0.7,0,0}{+0.77}  & \cellcolor{blue!5}\textcolor[rgb]{0.7,0,0}{+0.50}  & \cellcolor{blue!5}\textcolor[rgb]{0.7,0,0}{+1.11}  & \cellcolor{blue!5}\textcolor[rgb]{0.7,0,0}{+0.79}  & \cellcolor{blue!5}\textcolor[rgb]{0.7,0,0}{+0.83}  & \cellcolor{blue!5}\textcolor[rgb]{0.7,0,0}{+3.40}  & \cellcolor{blue!5}\textcolor[rgb]{0.7,0,0}{+0.47}  & \cellcolor{blue!5}\textcolor[rgb]{0.7,0,0}{+1.57}  \\
    \bottomrule
    \end{tabular}%
    }
    \caption{ Results of different approaches on varying datasets. We report the MUC, B$^3$, and CEAF$_{\phi_4}$ F1 scores (\%), their average (Avg. F1).  Of note, \textbf{Bold} and \underline{underline} highlight the best Avg.F1 within each approach (with varying $\tau_1/\tau_2$) and the best overall Avg.F1, respectively. $\Delta$ measures the average gain of our method over baselines: $\Delta = \frac{(a-b)+(a-c)}{2}$ ($a$: our method, $b$, $c$: baselines). ``Unbound'' means no constraint on cluster count for the linear classifier.}
  \label{tab:main_result}
  \vspace*{-2ex}
\end{table*}

\subsection{Main Results}
This section presents a systematic evaluation and comparative analysis of LRU, Dual-cache, and MEIC-DT across multiple datasets.
Note that LRU and Dual-cache only constrain the number of historical coreference clusters ($\tau_1$), whereas MEIC‑DT jointly constrains both the number of clusters ($\tau_1$) and the number of mentions per cluster ($\tau_2$) through the synergistic operation of its SAES and IRP modules. To comprehensively assess the performance of these methods, experiments are designed in two phases. First, incremental clustering is performed for $\tau_1 \in \{10,30,50,100,200\}$ by a linear classifier. Subsequently, under a Transformer architecture, experiments are conducted with $\tau_1/\tau_2 \in \{10/10, 10/20, 10/30, 30/30, 50/30\}$. 
In particular, since the IRP module is first introduced in this paper, we retain it while replacing the SAES strategy in MEIC‑DT with either LRU or Dual-cache, thereby constructing comparable hybrid frameworks.
Thus, in the experiments reported here, ``LRU+IRP'' and ``Dual‑cache+IRP'' actually correspond to LRU and Dual‑cache. The overall performance comparison of all methods is presented in Table~\ref{tab:main_result}.

\begin{figure*}[ht]
  \centering
  \includegraphics[width=1.0\linewidth]{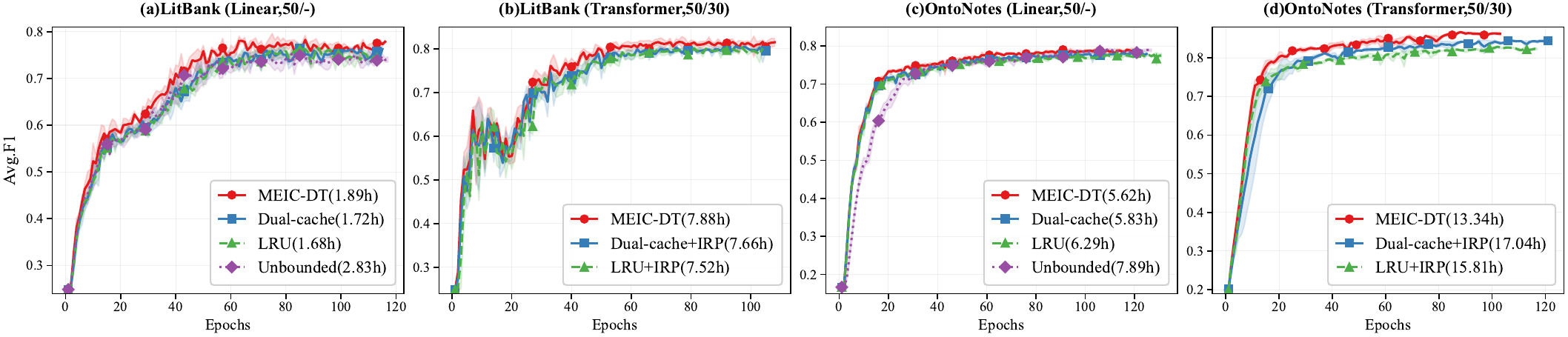}
  \caption{Learning curves and total training time. Results are shown for configurations with $\tau_1/\tau_2=50/-$ and $50/30$, across different datasets (LitBank, OntoNotes) and classifier backbones. } 
  \label{pic:speed_memory}
  \vspace*{-2ex}
\end{figure*}

\textbf{LitBank} Experiments on the LitBank dataset clearly reveal the combined impact of classifier architecture and constraint strategies. First, the Transformer Classifier consistently and significantly outperforms the Linear Classifier across all approaches. For instance, MEIC-DT achieves an Avg.F1 of 81.27\% under $\tau_1/\tau_2=50/30$, a substantial improvement over its linear counterpart (77.04\%), demonstrating stronger contextual modeling capability. Notably, when using the Linear Classifier, all approaches peak in performance at $\tau_1=50$ (with $\tau_2$ as -), while further relaxing constraints to $\tau_1=100$ or $200$ leads to a drop in Avg.F1. This highlights that classifiers with weaker contextual modeling struggle to effectively utilize excessively large memory capacity. Second, under different $\tau_1/\tau_2$ settings, the performance of all approaches follows a regular pattern: more relaxed constraints (e.g., $\tau_1=50, \tau_2=30$) generally yield the best results, indicating that providing moderate memory capacity is beneficial for this dataset. Finally, compared to baseline approaches, MEIC-DT significantly outperforms LRU+IRP (79.19\%) and Dual-cache+IRP (80.61\%) under identical constraint settings (e.g., $50/30$), with its $\Delta$ gains consistently positive. This validates the advantage of the synergistic optimization between SAES and IRP, particularly evident in key metrics such as B$^3$.

\textbf{OntoNotes} Results on OntoNotes further consolidate the effectiveness of our approach. The Transformer Classifier again demonstrates superior performance over the Linear Classifier, with MEIC-DT reaching a peak Avg.F1 of 85.05\% under the Transformer, higher than the 79.32\% of its linear version. Similarly, under the Linear Classifier, most approaches achieve their optimal Avg.F1 at $\tau_1=50$ (except LRU), with performance declining under more relaxed settings, further confirming the necessity of appropriate constraints. Regarding the tuning of $\tau_1/\tau_2$, an interesting observation is that under tighter memory constraints (e.g., $10/10$), the improvement of MEIC-DT over baselines (the $\Delta$ value) is particularly significant, indicating that its dual-threshold mechanism better maintains performance stability in resource-constrained scenarios. In direct comparison with existing baselines, MEIC-DT leads LRU+IRP and Dual-cache+IRP in Avg.F1 across almost all tested $\tau_1/\tau_2$ combinations, showing robust performance especially on the CEAF$_{\phi_4}$ metric, which highlights its improvements in clustering quality and consistency.

\textbf{WikiCoref} On the cross-domain generalization test set WikiCoref, this experiment evaluates the adaptability of models trained on OntoNotes. The results again confirm the effective generalization of the Transformer Classifier, with its performance across all approaches far surpassing that of the corresponding Linear Classifier versions. Even in out-of-domain testing, the Linear Classifier exhibits a similar trend, performing best at $\tau_1=50$ before degrading, underscoring the stability of the Transformer architecture in complex scenarios. Regarding constraint parameters, MEIC-DT achieves an optimal Avg.F1 of 72.63\% under $\tau_1/\tau_2=50/30$, suggesting that a balanced constraint configuration helps the model capture cross-domain coreference structures. More importantly, compared to the two baselines, MEIC-DT achieves the most significant lead on this out-of-domain test set (e.g., outperforming the best baseline, Dual-cache+IRP, by approximately 1.34 Avg.F1 points). Its $\Delta$ values are positive and substantial in most settings, strongly demonstrating that the mechanisms proposed in MEIC-DT possess superior domain generalization capability and robustness.

\begin{table}[htbp]
  \centering
  \resizebox{1.0\columnwidth}{!}{
    \begin{tabular}{lccc|ccc|cc}
    \toprule
    \multicolumn{1}{c}{\multirow{2}[2]{*}{$\tau_1$/$\tau_2$}} & \multicolumn{3}{c|}{\textbf{Litbank}} & \multicolumn{3}{c|}{\textbf{OntoNotes}} & \multicolumn{2}{c}{\textbf{WikiCoref}} \\
          & TAM & IAM & IAT & TAM & IAM & IAT & IAM & IAT \\
    \hline
    \rowcolor{mypink}\multicolumn{9}{c}{\textbf{Linear Classifier}} \\
    -/-   & 255.44 & 19.87 & 0.46  & 150.31 & 19.83 & 0.31  & 19.86 & 0.42 \\
    \hline
    10/-  & 72.22 & 19.15 & 0.62  & 52.86 & 19.14 & 0.37  & 19.15 & 0.46 \\
    \rowcolor{Gray}30/-  & 153.31 & 19.34 & 0.64  & 89.43 & 19.34 & 0.43  & 19.35 & 0.49 \\
    50/-  & 207.14 & 19.53 & 0.69  & 99.23 & 19.51 & 0.47  & 19.52 & 0.53 \\
    \rowcolor{Gray}100/- & 251.24 & 19.83 & 0.72  & 120.54 & 19.78 & 0.53  & 19.81 & 0.58 \\
    200/- & 255.44 & 19.87 & 0.46  & 147.37 & 19.81 & 0.55  & 19.84 & 0.59 \\
    \rowcolor{mypink}\multicolumn{9}{c}{\textbf{Transformer Classifier}} \\
    10/10 & 975.93 & 20.16 & 1.12  & 616.58 & 19.59 & 0.67  & 19.82 & 0.89 \\
    \rowcolor{Gray}10/20 & 1628.74 & 21.83 & 1.15  & 964.71 & 20.98 & 0.71  & 21.62 & 1.01 \\
    10/30 & 2118.95 & 23.43 & 1.22  & 1166.23 & 22.46 & 0.73  & 22.89 & 1.13 \\
    \rowcolor{Gray}30/30 & 7045.07 & 36.55 & 1.41  & 3046.73 & 31.93 & 0.77  & 36.31 & 1.29 \\
    50/30 & 14581.14 & 38.68 & 1.81  & 7635.27 & 38.33 & 0.80  & 38.49 & 1.68 \\
    \bottomrule
    \end{tabular}%
    }
    \caption{Average resource consumption under identical settings. Note: Metrics are averaged across different approaches for each fixed configuration of dataset, classifier, and $\tau_1/\tau_2$. TAM, IAM, and IAT denote Training Average Memory (MB), Inference Average Memory (MB), and Inference Average Time (s), respectively.}
  \label{tab:speed_memory}%
  \vspace*{-2ex}
\end{table}%

\subsection{Speed and Memory Usage}
Our experiments reveal that under identical configurations of dataset, classifier, and $\tau_1/\tau_2$ constraints, different approaches exhibit negligible differences in terms of Training Average Memory (TAM), Inference Average Memory (IAM), and Inference Average Time (IAT). Therefore, we report the averaged values of these metrics under each such fixed setting in Table~\ref{tab:speed_memory}. In contrast, the training dynamics and time consumption vary significantly across approaches. Figure~\ref{pic:speed_memory} presents the learning curves and total training time for various approaches under the $\tau_1/\tau_2$ settings of $50/-$ and $50/30$, across different datasets and classifiers.

As shown in Table \ref{tab:speed_memory}, the Transformer Classifier incurs significantly higher TAM than the Linear Classifier due to its more complex architecture. However, the gap in IAM and IAT between the two is relatively small, indicating that the constraint configuration ($\tau_1/\tau_2$) is the dominant factor for inference cost. As $\tau_1/\tau_2$ increases, both IAM and IAT rise monotonically across all configurations. Notably, under the Transformer architecture, memory consumption surges sharply from $30/30$ to $50/30$. This trend directly aligns with the performance peak observed at $50/30$ (see Table \ref{tab:main_result}), clearly revealing the inherent trade-off between performance and efficiency.
Figure \ref{pic:speed_memory} provides a deeper analysis of the training dynamics. On LitBank, MEIC-DT slightly increases training time compared to baselines, yet its learning curve consistently dominates, indicating higher convergence efficiency. On OntoNotes, MEIC-DT demonstrates a more pronounced advantage: it not only requires substantially less training time (e.g., only 13.34 hours with the Transformer backbone, significantly lower than the baselines' 17.04 and 15.81 hours) and fewer epochs but also exhibits a learning curve that is superior throughout or in later stages. Together, these results demonstrate that the dual-threshold mechanism of MEIC-DT not only enhances final performance (see Table \ref{tab:main_result}) but also accelerates model convergence through a more efficient and stable optimization process, achieving a dual advantage in both effectiveness and efficiency.

\subsection{Ablation Study}

\textbf{The Efficacy of SAES.} We look into  the efficacy of SAES in Litbank and OntoNotes using Transformer classifier with $\tau_1/\tau_2=50/30$. As detailed in Section~\ref{sec:saes}, MEIC-DT employs distinct eviction criteria for training and inference, termed SAES$_{\text{train}}$ and SAES$_{\text{inf}}$.
To ablate the contribution of each stage, we construct hybrid variants by replacing the corresponding SAES module with either LRU or Dual‑cache during training or inference. We report results in Table \ref{tab:impact_saes}.
As shown in Table~\ref{tab:impact_saes}, employing the SAES$_{\text{train}}$/SAES$_{\text{inf}}$ strategy yields the highest Avg.F1 on both LitBank and OntoNotes. Replacing SAES$_{\text{inf}}$ with LRU during inference alone causes a marked performance drop (e.g., from 81.27\% to 79.43\% on LitBank), underscoring the effectiveness of SAES$_{\text{inf}}$ in preserving critical clusters via mention activity ($em_j/age_j$) in the absence of gold annotations. Similarly, substituting SAES$_{\text{train}}$ with LRU or Dual-cache in training also degrades results, validating the superiority of its eviction criterion based on gold remaining mentions ($rm_j$). These results collectively demonstrate that the stage-specific strategies of SAES---tailored to the distinct information available during training and inference---are both indispensable, and their synergy is crucial to MEIC-DT's leading performance.

\begin{table}[htbp]
  \centering
  \resizebox{1.0\columnwidth}{!}{
    \begin{tabular}{ccccc}
    \toprule
    \textbf{Train}/\textbf{Inference} & \textbf{MUC}   & \textbf{B}$^3$   & \textbf{CEAF}$_{\phi_4}$ & \textbf{\textcolor[rgb]{0.8,0,0}{Avg.F1}} \\
    \rowcolor{mypink}\multicolumn{5}{c}{\textbf{Litbank}} \\
    SAES$_{\text{train}}$/SAES$_{\text{inf}}$ & 87.85$_{\pm0.28}$ & 78.53$_{\pm0.24}$ & 77.42$_{\pm0.03}$ & \textbf{81.27$_{\pm0.41}$} \\
    \hline
    SAES$_{\text{train}}$/LRU & 86.87$_{\pm0.54}$ & 75.42$_{\pm0.31}$ & 76.01$_{\pm0.22}$ & 79.43$_{\pm0.25}$ \\
    SAES$_{\text{train}}$/Dual-cache & 87.56$_{\pm0.46}$ & 77.08$_{\pm0.28}$ & 77.34$_{\pm0.62}$ & 80.66$_{\pm0.39}$ \\
    \hdashline
    LRU/SAES$_{\text{inf}}$ & 86.17$_{\pm0.51}$ & 75.63$_{\pm0.11}$ & 75.06$_{\pm0.38}$ & 79.15$_{\pm0.21}$ \\
    \rowcolor{Gray}Dual-cache/SAES$_{\text{inf}}$ & 87.02$_{\pm0.38}$ & 76.99$_{\pm0.26}$ & 76.89$_{\pm0.14}$ & 80.30$_{\pm0.27}$ \\
    SAES$_{\text{inf}}$/SAES$_{\text{inf}}$ & 87.45$_{\pm0.42}$ & 77.29$_{\pm0.48}$ & 77.38$_{\pm0.38}$ & 80.71$_{\pm0.67}$ \\
    \rowcolor{mypink}\multicolumn{5}{c}{\textbf{OntoNotes}} \\
    SAES$_{\text{train}}$/SAES$_{\text{inf}}$ & 90.53$_{\pm0.43}$ & 84.43$_{\pm0.42}$ & 80.19$_{\pm0.49}$ & \textbf{85.05$_{\pm0.67}$} \\
    \hline
    SAES$_{\text{train}}$/LRU & 89.29$_{\pm0.54}$ & 82.57$_{\pm0.14}$ & 79.25$_{\pm0.35}$ & 83.71$_{\pm0.16}$ \\
    SAES$_{\text{train}}$/Dual-cache & 90.31$_{\pm0.32}$ & 83.46$_{\pm0.24}$ & 80.69$_{\pm0.14}$ & 84.82$_{\pm0.52}$ \\
    \hdashline
    LRU/SAES$_{\text{inf}}$ & 89.48$_{\pm0.17}$ & 82.52$_{\pm0.26}$ & 79.41$_{\pm0.33}$ & 83.80$_{\pm0.29}$ \\
    \rowcolor{Gray}Dual-cache/SAES$_{\text{inf}}$ & 90.33$_{\pm0.58}$ & 82.35$_{\pm0.47}$ & 80.89$_{\pm0.34}$ & 84.86$_{\pm0.71}$ \\
    SAES$_{\text{inf}}$/SAES$_{\text{inf}}$ & 90.28$_{\pm0.61}$ & 83.45$_{\pm0.57}$ & 79.96$_{\pm0.38}$ & 84.89$_{\pm0.55}$ \\
    \bottomrule
    \end{tabular}%
    }
    \caption{The Impact (\%) of the SAES Module in MEIC-DT. Evaluation on LitBank and OntoNotes with the Transformer classifier under $\tau_1/\tau_2=50/30$.}
  \label{tab:impact_saes}%
  \vspace*{-2ex}
\end{table}%

\begin{table}[htbp]
  \centering
  \resizebox{1.0\columnwidth}{!}{
    \begin{tabular}{ccccc}
    \toprule
    \textbf{IRP}   & \textbf{MUC}   & \textbf{B}$^3$   & \textbf{CEAF}$_{\phi_4}$ & \textbf{\textcolor[rgb]{0.8,0,0}{Avg.F1}} \\
    \rowcolor{mypink}\multicolumn{5}{c}{\textbf{Litbank}} \\
    Group-based Sampling (ours) & 87.85$_{\pm0.28}$ & 78.53$_{\pm0.24}$ & 77.42$_{\pm0.03}$ & \textbf{81.27$_{\pm0.41}$} \\
    Random Sampling & 87.12$_{\pm0.15}$ & 76.98$_{\pm0.27}$ & 77.01$_{\pm0.13}$ & 80.37$_{\pm0.22}$ \\
    \rowcolor{mypink}\multicolumn{5}{c}{\textbf{OntoNotes}} \\
    Group-based Sampling (ours) & 90.53$_{\pm0.43}$ & 84.43$_{\pm0.42}$ & 80.19$_{\pm0.49}$ & \textbf{85.05$_{\pm0.67}$} \\
    Random Sampling & 89.08$_{\pm0.22}$ & 82.33$_{\pm0.54}$ & 80.04$_{\pm0.19}$ & 83.82$_{\pm0.37}$ \\
    \bottomrule
    \end{tabular}%
    }
    \caption{The utility (\%) of the IRP Module in MEIC-DT. Evaluation on LitBank and OntoNotes with the Transformer classifier under $\tau_1/\tau_2=50/30$.}
  \label{tab:utility_irp}%
\end{table}%

\begin{figure}[ht]
  \centering
  \includegraphics[width=1.0\linewidth]{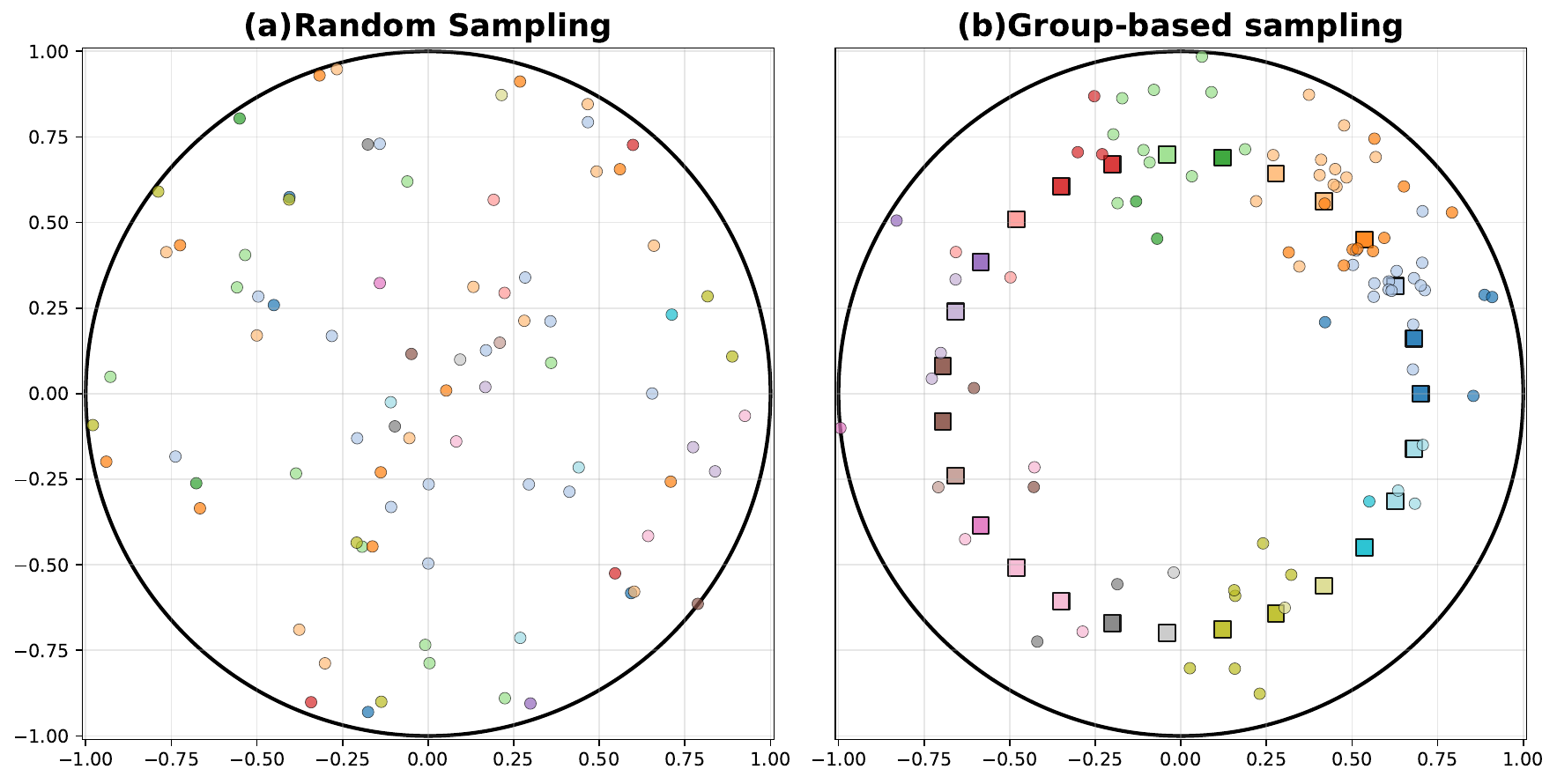}
  \caption{An example of semantic space distributions for different sampling strategies on OntoNotes.} 
  \label{pic2_fig:}
\end{figure}

\textbf{The utility of IRP.} 
We conduct an ablation study to assess the utility of the IRP module in MEIC-DT on OntoNotes and LitBank, using the Transformer classifier under $\tau_1/\tau_2=50/30$. As introduced in Section~\ref{sec:irp}, IRP is a core innovation that employs \textbf{\textit{group-based sampling}} by leveraging mention textual relations to select the most representative mentions per cluster. To validate this strategy, we compare it against a \textbf{\textit{random sampling}} baseline, which selects mentions uniformly at random from each cluster without grouping. Results are presented in Table~\ref{tab:utility_irp}. From \ref{tab:utility_irp}, the group-based sampling significantly outperforms random sampling on both LitBank and OntoNotes (e.g., achieving a +1.23 gain in Avg.F1 on OntoNotes). This confirms that IRP can more effectively preserve the core semantic information of each cluster through semantic-relation-based grouping and stratified sampling. The semantic space visualization in Fig.~\ref{pic2_fig:} further reveals that the mentions selected by group-based sampling exhibit a more reasonable and representative distribution, which directly explains its superiority in maintaining CR performance and validates the key role of IRP as a structured information compression strategy.

\section{Related Works}
\label{Related_works:app}
Coreference Resolution (CR) has long been a crucial task in the field of NLP. The seminal work by~\cite{lee2017end} introduced the first end-to-end supervised neural method, i.e., the Coarse-to-Fine model, establishing the detect-then-cluster pipeline. 
Shortly thereafter, a panoply of efforts have focused on improving this method's training efficiency, memory consumption, and coreference performance. 
For example, some methods~\cite{dobrovolskii-2021-word, si-etal-2022-scl, kantor2019coreference, xu-choi-2020-revealing, wu-etal-2020-corefqa, kirstain2021coreference,lai2022end, otmazgin2022f,otmazgin2022lingmess,d2023caw,sundar-etal-2024-major} retained the detect-then-cluster pipeline and leveraged pre-trained language models (e.g., BERT~\cite{devlin-etal-2019-bert}, SpanBERT~\cite{joshi2020spanbert}, Longformer~\cite{beltagy2020longformer}) for document encoding, significantly boosting performance. However, they still grapple with substantial memory overhead. 
Also, generative methods~\cite{liu-etal-2022-autoregressive, bohnet-etal-2023-coreference, zhang-etal-2023-seq2seq}, centered around large sequence-to-sequence architectures, once dominated the pursuit of high coreference performance. Nevertheless, their prohibitively expensive training costs and high inference latency render their deployment infeasible.

With super-sized training corpora and computational cluster resources, LLMs have demonstrated powerful reasoning capabilities, thus enabling SOTA performance in a wide range of natural language tasks~\cite{achiam2023gpt, liu2024deepseek,qwen3technicalreport, luo2024let, tan2024llm, yu2024automated,si2025goal,wang2025document, si2025teaching,luo2025gltw,si2025aligning,si2024gateau, bai2025sis}. 
However, their performance in CR has yet to surpass mainstream supervised neural methods~\cite{le2023large, gan-etal-2024-assessing}. This limitation stems from the ``large but not precise'' nature of LLMs, which hinders high-accuracy mention extraction~\cite{liu2024bridging}. Intriguingly, under ideal conditions where gold mentions are provided, LLMs leverage their strong reasoning capabilities to achieve CR performance that matches or even exceeds that of supervised neural methods~\cite{le2023large}. Recently, other attempts to incorporate LLMs into CR either rely on overly idealistic assumptions~\cite{sundar-etal-2024-major} or entail prohibitive computational costs to train multiple LLMs~\cite{zhu-etal-2025-llmlink}.


Of note, under \textit{supervised neural methods}, there exists another line of works~\cite{xia-etal-2020-incremental, toshniwal-etal-2020-learning, guo-etal-2023-dual, martinelli-etal-2024-maverick} to drew inspiration from human cognitive incremental processing mechanisms for long-text scenarios. During the mention clustering, they progressively evaluate the association between candidate mentions and existing coreference clusters using linear classifiers or lightweight Transformer architectures to determine coreference links. 
However, despite significant progress in CR, they still face a critical challenge in balancing efficiency and performance during mention clustering. 
For example, conventional incremental clustering approaches~\cite{xia-etal-2020-incremental, toshniwal-etal-2020-learning, guo-etal-2023-dual} typically employ a fixed-size cache and a linear classifier to associate candidate mentions with coreference clusters (each represented as a single vector) tracked by the cache. 
These approaches decide whether a mention should join an existing cluster or form a new one. 
Upon reaching capacity, a predefined rule-based eviction strategy---LRU (\textit{Least Recently Used}) or a dual-cache mechanism combining LRU and LFU (\textit{Least Frequently Used})---removes certain entities to update the cache dynamically.  
While computationally efficient, such approaches often fail to capture the complex nonlinear dependencies between mentions and clusters, and yield limited semantic representations due to excessive intra-cluster compression. 

Moreover, these strategies either rely on a single heuristic (e.g., LRU) that may discard essential clusters, or employ a dual-cache mechanism that increases the complexity of cache management. 
Crucially, the same eviction strategy is used across both training and inference. This does not account for the inherent differences in information availability between the two phases and thus constrains the model's overall effectiveness.
Recently, \citet{martinelli-etal-2024-maverick} introduced a lightweight Transformer model. 
Its input is formed by concatenating the representation of a candidate mention with representations of all mentions from historical coreference clusters, effectively modeling the global intra-cluster structure and significantly improving clustering accuracy.
However, a major drawback is that the input size grows polynomially with the number of clusters and the number of mentions per cluster, causing memory usage and computational latency to surge during training, which severely limits its practicality and scalability given academic computational resources. 

\section{Conclusion}
This paper presents MEIC-DT, a memory-efficient incremental clustering model for CR that operates under strict dual-threshold memory constraints. The core of our approach is a dual-threshold constraint mechanism, instantiated through two novel components: a Statistics-Aware Eviction Strategy for phase-adaptive cache management, and an Internal Regularization Policy for semantically representative cluster condensation. Extensive experiments show that MEIC-DT achieves a superior balance between performance and memory efficiency compared to existing baselines, validating the effectiveness of its synergistic design in managing the inherent trade-offs for long-text processing.

\section*{Limitations}
While our method demonstrates significant advances in memory-efficient incremental clustering for coreference resolution, several limitations warrant discussion. \textbf{\textit{First}}, in the current era dominated by large language models (LLMs), our work—focusing on supervised neural methods—might appear to swim against the tide. However, we note that recent studies~\cite{liu2024bridging} have shown high-performing supervised coreference resolvers can effectively enhance LLMs' capabilities in downstream tasks such as generation and question answering, which presents a clear and valuable application niche for our efficient model. \textbf{\textit{Second}}, our experiments were conducted under constrained computational budgets. We acknowledge that with more generous resources (e.g., larger $\tau_1/\tau_2$ or model capacity), performance could potentially be further improved. \textbf{\textit{Finally}}, we believe more effective memory constraint strategies may exist. Future work could explore adaptive threshold learning or reinforcement learning-based policies for dynamic memory allocation.

\section*{Acknowledgments}
This work is supported by the National Natural Science Foundation of China (No.T2341003), Beijing Municipal Science and Technology Plan Project (Z241100001324025) and a grant from the Guoqiang Institute, Tsinghua University.


\bibliography{custom}

\clearpage
\appendix

\section*{Appendix}
\label{sec:appendix}

\noindent This appendix is organized as follows.  

\begin{itemize}
    \item In Section~\ref{app:code_IRP}, we present the detailed algorithmic pseudocode for Internal Regularization Policy.
    \item In Section~\ref{app:full_pronouns_list}, we list all the pronouns.
    \item In Section~\ref{app:com_exp_set}, we provide a comprehensive description of the experimental setup, including the datasets employed, baseline methods compared, evaluation metrics utilized, and detailed implementation configurations.
    \item In Section~\ref{app:add_exp_res}, we present additional experimental results.
\end{itemize}




\section{Pseudocode of Internal Regularization Policy}
\label{app:code_IRP}
In this section, we present the detailed algorithmic pseudocode for Internal Regularization Policy, as shown in Algorithms~\ref{alg:irp}-\ref{alg:sampling}.

\begin{algorithm}
\caption{Internal Regularization Policy (\textbf{IRP})}
\label{alg:irp}
\begin{algorithmic}[1]
\REQUIRE Cluster set $\mathcal{C} = \{c_1, c_2, \dots, c_{\tau_1}\}$ where each $c_i = [(t_{i,s}^1, t_{i,e}^1), (t_{i,s}^2, t_{i,e}^2), \dots]$ represents mention spans, original word text $words$
\REQUIRE Maximum cluster size threshold $\tau_2$
\ENSURE Compressed cluster set $\mathcal{C}'$ where each cluster has at most $\tau_2$ mentions

\FOR{each cluster $c_i \in \mathcal{C}$}
    \IF{$|c_i| > \tau_2$}
        \STATE Sort $c_i$ in ascending order, i.e., $\text{sort}(c_i)$
        \STATE Initialize compressed cluster: $c_i' \leftarrow [c_i[0]]$ \COMMENT{Preserve first mention}
        \STATE $\hat{c}_i \leftarrow c_i[1:-1]$ \COMMENT{Exclude first and last mentions}
        \STATE $\hat{\mathcal{C}}_i^{grouped} \leftarrow $ $\text{\textbf{SG}}(words, $ $\hat{c}_i)$
        \STATE $\hat{c}_i^{sampled} \leftarrow \text{\textbf{GS}}(\hat{\mathcal{C}}_i^{grouped}, \tau_2 - 2)$
        \STATE $c_i'.extend(\hat{c}_i^{sampled})$
        \STATE $c_i'.append(c_i[-1])$ \COMMENT{Preserve last mention}
        \STATE $\mathcal{C}' \leftarrow \mathcal{C}' \cup \{\text{sort}(c_i')\}$
    \ELSE
        \STATE $\mathcal{C}' \leftarrow \mathcal{C}' \cup \{c_i\}$ \COMMENT{Keep original cluster}
    \ENDIF
\ENDFOR
\RETURN $\mathcal{C}'$
\end{algorithmic}
\end{algorithm}

\begin{algorithm*}
\caption{Semantic Grouping (\textbf{SG})}
\label{alg:grouping}
\begin{algorithmic}[1]
\REQUIRE $words$, $\hat{c}$: list of mention spans $[(t_{1,s}, t_{1,e}), (t_{2,s}, t_{2,e}), \dots, (t_{|\hat{c}|-1,s}, t_{|\hat{c}|-1,e})]$, and $\mathcal{P}$: pronoun lexicon (see Appendix~\ref{app:full_pronouns_list})
\ENSURE $\hat{\mathcal{C}}^{grouped}$: semantically grouped mention spans

\STATE $mentions \leftarrow [\,]$ \COMMENT{Store mention texts}
\FOR{each $(t_{s}, t_{e}) \in \hat{c}$}
    \STATE $span=[]$
    \FOR {$idx \in [t_{s}, t_{e}]$}
        \STATE $span.append(words[idx].lower())$
    \ENDFOR
    \STATE $mentions.append(\text{concat}(span))$
\ENDFOR

\STATE $n \leftarrow |\hat{c}|$
\STATE $uf \leftarrow \text{UnionFind}(n)$ \COMMENT{Initialize Union-Find data structure}

\FOR{$i = 0$ to $n-1$}
    \FOR{$j = i$ to $n-1$}
        \IF{$mentions[i] \in \mathcal{P} \land mentions[j] \in \mathcal{P}$} 
            \IF{$mentions[i] = mentions[j]$}
                \STATE $uf.union(i, j)$
            \ENDIF
        \ELSIF{$mentions[i] \notin \mathcal{P} \land mentions[j] \notin \mathcal{P}$}
            \IF{$mentions[i] = mentions[j] \lor mentions[i] \subseteq mentions[j] \lor mentions[j] \subseteq mentions[i]$}
                \STATE $uf.union(i, j)$
            \ENDIF
        \ENDIF
    \ENDFOR
\ENDFOR

\STATE Initialize $groups \leftarrow \{\}$
\FOR{$i = 0$ to $n-1$}
    \STATE $root \leftarrow uf.find(i)$
    \STATE $groups[root].append(i)$
\ENDFOR

\STATE $\hat{\mathcal{C}}^{grouped} \leftarrow [\,]$
\FOR{each $group \in groups.values()$}
    \STATE $cluster\_group=[]$
    \FOR {$i \in group$}
        \STATE $cluster\_group.append(cluster\_idxs[i])$
    \ENDFOR
    \STATE $\hat{\mathcal{C}}^{grouped}.append(cluster\_group)$
\ENDFOR

\RETURN $\hat{\mathcal{C}}^{grouped}$
\end{algorithmic}
\end{algorithm*}

\begin{algorithm*}
\caption{Group-based Sampling (\textbf{GS})}
\label{alg:sampling}
\begin{algorithmic}[1]
\REQUIRE $\hat{\mathcal{C}}_i^{grouped} = [G_1, G_2, \dots, G_m]$: grouped mention spans, and $k$: number of mentions to sample
\ENSURE $\hat{c}^{sampled}$: selected mention spans

\STATE $m \leftarrow |\hat{\mathcal{C}}^{grouped}|$ \COMMENT{Number of groups}
\STATE $n_{list} \leftarrow [|G_i| \text{ for } G_i \in \hat{\mathcal{C}}^{grouped}]$ \COMMENT{Sizes of each group}
\STATE $N \leftarrow \sum n_{list}$ \COMMENT{Total mentions}

\IF{$k > N$}
    \STATE \textbf{error} "Cannot sample more mentions than available"
\ENDIF

\IF{$k < m$} 
    \STATE \COMMENT{More groups than needed samples}
    \STATE Sort groups in $\hat{\mathcal{C}}^{grouped}$ by size in descending order
    \STATE $samples \leftarrow [\,]$
    \FOR{$i = 0$ to $k-1$}
        \STATE Randomly select one mention from $G_i$
        \STATE $samples.append(selected\_mention)$
    \ENDFOR
    \RETURN $samples$
\ENDIF

\COMMENT{Case when $k \geq m$: allocate at least one sample per group}
\STATE $s \leftarrow [1, 1, \dots, 1]$ \COMMENT{Initialize with one sample per group}
\STATE $r \leftarrow k - m$ \COMMENT{Remaining samples to allocate}

\IF{$r > 0$}
    \STATE $capacity_i \leftarrow [n_i - 1 \text{ for } n_i \in n_{list}]$ \COMMENT{Remaining capacity per group}
    \STATE $total\_capacity \leftarrow \sum capacity_i$
    \IF{$r > total\_capacity$}
        \STATE $r \leftarrow total\_capacity$
    \ENDIF
    
    \WHILE{$r > 0$}
        \STATE Find group $j$ with maximum $capacity_j > 0$
        \STATE $s[j] \leftarrow s[j] + 1$
        \STATE $capacity_j \leftarrow capacity_j - 1$
        \STATE $r \leftarrow r - 1$
    \ENDWHILE
\ENDIF

\STATE $\hat{c}^{sampled} \leftarrow [\,]$
\FOR{$i = 0$ to $m-1$}
    \STATE Randomly select $s[i]$ mentions from group $G_i$
    \STATE $\hat{c}^{sampled}.extend(selected\_mentions)$
\ENDFOR
\RETURN $\hat{c}^{sampled}$
\end{algorithmic}
\end{algorithm*}

\section{Full Pronouns List}
\label{app:full_pronouns_list}
We list all the pronouns in Table~\ref{tab:full_pronouns_list}. 

\begin{table}[htbp]
  \centering
  \resizebox{1.0\columnwidth}{!}{
    \begin{tabular}{p{20.455em}}
    \toprule
    i , me , my , mine , myself , you , your , yours , yourself , yourselves , he , him , his , himself , she , her , hers , herself , it , its , itself , we , us , our , ours , ourselves , they , them , their , themselves , that , this  \\
    \bottomrule
    \end{tabular}%
    }
    \caption{Full pronouns list. Note that in the specific experimental process, ``it'' should be removed from full pronouns list because ``it'' may lack referential significance. }
  \label{tab:full_pronouns_list}%
\end{table}%

\section{Complete Experimental Settings}
\label{app:com_exp_set}
\textbf{Datasets and Baselines.} We train and evaluate the performance of our proposed methods on two widely used CR datasets: OntoNotes~\cite{pradhan-etal-2012-conll} and LitBank~\cite{bamman-etal-2020-annotated}. 
Specifically, OntoNotes, originating from the CoNLL-2012 shared task, serves as the de facto standard for evaluating CR methods. It encompasses seven text genres, including full-length documents such as newswire articles, broadcast news, magazines, web text, and testament passages, as well as multi-speaker transcripts like broadcast conversations and telephone conversations. LitBank is frequently employed for evaluating CR on long documents and comprises $100$ literary works. 
Note that OntoNotes and LitBank follow different annotation guidelines: OntoNotes does not annotate singleton clusters (i.e., single-mention clusters), whereas LitBank provides such annotations. 
To assess the generalization capability of methods, models trained on OntoNotes are also tested on the out-of-domain dataset WikiCoref~\cite{ghaddar-langlais-2016-wikicoref}. 
WikiCoref only contains $30$ Wikipedia articles as its test set, with some texts reaching lengths of up to $9869$ words. 
We detail these datasets in Table~\ref{tab:Dataset_statistics_app}.

Meanwhile, to gauge the effectiveness of MEIC-DT, we select the most relevant baselines in the main paper: LRU~\cite{xia-etal-2020-incremental, toshniwal-etal-2020-learning} and Dual-cache~\cite{guo-etal-2023-dual}.
We follow~\citet{luo2025imcoref} in adopting Imcoref as Mention Extractor and DeBERTa$_{\text{large}}$~\cite{he2021deberta} as Text Encoder~(see Fig.~\ref{Framework_fig:}). Furthermore, we conduct coreference clustering experiments adopting two distinct classifiers---a linear classifier~\cite{xia-etal-2020-incremental, toshniwal-etal-2020-learning,guo-etal-2023-dual} and a lightweight Transformer classifier~\cite{martinelli-etal-2024-maverick}---to thoroughly evaluate the effectiveness of MEIC-DT.
As outlined in Related Works (see Appendix~\ref{Related_works:app}), numerous CR approaches exist. Accordingly, for more comparison results, we report additional experimental results in Tables~\ref{tab:Per_com_litbank}-\ref{tab:Per_com_ontonotes} from Appendix~\ref{app:add_exp_res}.

\begin{table}[htbp]
  \centering
  \resizebox{1.0\columnwidth}{!}{
    \begin{tabular}{l|ccccc}
    \hline
    Datasets & \#Train & \#Val & \#Test & Avg. W & Avg. M  \\
    \hline
    OntoNotes & 2802  & 343   & 348   & 467   & 56      \\
    LitBank & 80    & 10    & 10    & 2105  & 291     \\
    WikiCoref & -     & -     & 30    & 1996  & 230    \\
    \hline
    \end{tabular}%
    }
    \caption{Dataset statistics: number of documents in each dataset split (\#Train/Val/Test), average number of words (i.e., Avg. W) and mentions (i.e., Avg. M) per document.}
  \label{tab:Dataset_statistics_app}%
\end{table}%

\textbf{Metrics.}
In our experiments, we evaluate performance using the F1 scores of the MUC~\cite{vilain1995model}, B$^3$~\cite{bagga1998algorithms}, and CEAF$_{\phi_4}$~\cite{luo2005coreference} metrics. The overall CoNLL-F1 score (abbreviated as Avg.F1) is the average of these three F1 scores. To ensure reliability, we report the average for each experiment over $3$ different random seeds.

\textbf{Configurations.}
In all experiments, unless otherwise specified, we adopt the default settings of $\tau_1=50$ and $\tau_2=30$ for the Transformer classifier, while maintaining $\tau_1=50$ for the linear classifier. When implementing the linear classifier, we follow~\citet{xia-etal-2020-incremental,toshniwal-etal-2020-learning,guo-etal-2023-dual} by representing each historical coreference cluster as a vector representation. This formulation eliminates the necessity of additional parameters beyond the threshold $\tau_1$. We set $\delta=1\times 10^{-5}$ and employ the Adafactor optimizer~\cite{shazeer2018adafactor} for model training, with learning rates configured as 2e-5 for DeBERTa$_{\text{large}}$ and 3e-4 for the remaining model layers.
All experiments are implemented using the PyTorch-Lightning framework. Each run is executed on a single RTX 4090 GPU with 24GB of VRAM.
We have eight such GPUs available, allowing multiple experiments to be conducted concurrently. For fairness, neither data nor model parallelism is employed in any experiment.
During training, we accumulate gradients every four steps and set the gradient clipping threshold at $1.0$. A linear learning rate scheduler is adopted, incorporating a warm-up phase covering $10\%$ of all training steps. To monitor performance, a validation evaluation is performed every one epoch. The final model is selected based on Avg.F1 in the validation set, with an early stopping patience of $30$.

\section{Additional Experimental Results}
\label{app:add_exp_res}
In this section, we report additional experimental results, as detailed in Tables~\ref{tab:Per_com_litbank}-\ref{tab:Per_com_ontonotes}. 

\begin{table}[htbp]
  \centering
  \resizebox{1.0\columnwidth}{!}{
    \begin{tabular}{lccc|c}
    \toprule
    \multicolumn{1}{c}{Methods} & MUC   & B$^3$ & CEAF$_{\phi_4}$ & \textcolor[rgb]{0.8,0,0}{\textbf{Avg.F1}} \\
    \hline
    longdoc~\cite{toshniwal2021generalization} & 88.2  & 75.9  & 65.5  & 76.5  \\
    Dual-cache~\cite{guo-etal-2023-dual} & 88.2  & 79.2  & 71.0  & 79.5  \\
    Maverick$_{\text{mes}}$~\cite{martinelli-etal-2024-maverick} & 86.2  & 78.4  & 69.6  & 78.1  \\
    Maverick$_{\text{incr}}$~\cite{martinelli-etal-2024-maverick} & 86.5  & 78.8  & 69.8  & \textbf{78.3}  \\
    ImCoref$_{\text{mes}}$~\cite{luo2025imcoref} & 87.9  & 79.5  & 71.6  & \textbf{79.7}  \\
    MEIC-DT(\textbf{ours}) & 87.9  & 78.5  & 77.4  & \textbf{81.3}  \\
    \hline
    seq2seq~\cite{zhang-etal-2023-seq2seq} & -     & -     & -     & 77.3 \\
    \hline
    GPT-3.5-turbo~\cite{le2023large} & -     & -     & -     & 75.3  \\
    LLMLink~\cite{zhu-etal-2025-llmlink} & -     & -     & -     & 81.5$^\star$  \\
    \hline
    ImCoref-CeS$_{\text{gpt4}}$~\cite{luo2025imcoref} & 88.8  & 81.5  & 72.9  & 81.1  \\
    \bottomrule
    \end{tabular}%
    }
    \caption{Performance comparison (\%) of different methods on LitBank. Note that ($\star$)  indicates that it is required to train multiple LLMs to perform coreference task.}
  \label{tab:Per_com_litbank}%
  \vspace*{-2ex}
\end{table}%

\begin{table*}[t]
  \centering
  \resizebox{1.95\columnwidth}{!}{
    \begin{tabular}{llccccccl}
    \toprule
    \multicolumn{1}{c}{\multirow{2}[4]{*}{Methods}} & \multicolumn{1}{c}{\multirow{2}[4]{*}{Base Encoders}} & \multirow{2}[4]{*}{MUC  } & \multirow{2}[4]{*}{B$^3$} & \multirow{2}[4]{*}{CEAF$_{\phi_4}$} & \multirow{2}[4]{*}{\textcolor[rgb]{0.8,0,0}{\textbf{Avg.F1}}} & \multirow{2}[4]{*}{Params} & \multicolumn{2}{c}{Training} \\
\cline{8-9}          &       &       &       &       &       &       & Time  & \multicolumn{1}{c}{Hardware} \\
    \hline
     \rowcolor{mypink}\multicolumn{9}{l}{\textbf{Supervised neural methods}} \\
    c2f-coref~\cite{joshi2020spanbert} & SpanBERT$_\text{large}$ & 85.3  & 78.1  & 75.3  & 79.6  & 370M  & -     & \multicolumn{1}{r}{1$\times$ 32G} \\
    Icoref~\cite{xia-etal-2020-incremental} & SpanBERT$_\text{large}$ & 85.3  & 77.8  & 75.2  & 79.4  & 377M  & 40h   & \multicolumn{1}{r}{1$\times$1080TI-12G} \\
    CorefQA~\cite{wu-etal-2020-corefqa} & SpanBERT$_\text{large}$ & 88.0  & 82.2  & 79.1  & 83.1$^\diamond$  & 740M  & -     & \multicolumn{1}{r}{1$\times$TPUv3-128G} \\
    s2e-coref~\cite{kirstain2021coreference} & LongFormer$_\text{large}$ & 85.8  & 79.1  & 76.1  & 80.3  & 494M  & -     & \multicolumn{1}{r}{1$\times$ 32G} \\
    longdoc~\cite{toshniwal2021generalization} & LongFormer$_\text{large}$ & 85.3  & 78.0  & 75.3  & 79.6  & 471M  & 16h   & \multicolumn{1}{r}{1$\times$ A6000-48G} \\
    wl-coref~\cite{dobrovolskii-2021-word} & RoBERTa$_\text{large}$ & 86.3  & 79.9  & 76.6  & 81.0  & 360M  & 5h    & \multicolumn{1}{r}{1$\times$RTX8000-48G} \\
    f-coref~\cite{otmazgin2022f} & DistilRoBERTa & 84.4  & 76.6  & 74.5  & 78.5$^\diamond$  & 91M   & -     & \multicolumn{1}{r}{1$\times$V100-32G} \\
    LingMess~\cite{otmazgin2022lingmess} & LongFormer$_\text{large}$ & 86.6  & 80.5  & 77.3  & 81.4  & 590M  & 23h   & \multicolumn{1}{r}{1$\times$V100-32G} \\
    Dual-cache~\cite{guo-etal-2023-dual} & LongFormer$_\text{large}$ & 86.3  & 80.3  & 76.8  & 81.1  & 471M  & -     & \multicolumn{1}{r}{1$\times$T4-16G} \\
    Maverick$_\text{{mes}}$~\cite{martinelli-etal-2024-maverick} & DeBERTa$_\text{large}$ & 88.0  & 82.8  & 79.9  & 83.6  & 504M  & 14h   & \multicolumn{1}{r}{1$\times$RTX4090-24G} \\
    Maverick$_\text{{incr}}$~\cite{martinelli-etal-2024-maverick} & DeBERTa$_\text{large}$ & 87.9  & 82.7  & 79.8  & 83.5  & 452M  & 29h   & \multicolumn{1}{r}{1$\times$RTX4090-24G} \\
    ImCoref$_\text{{mes}}$~\cite{luo2025imcoref} & DeBERTa$_\text{large}$ & 88.2  & 83.6  & 81.0  & 84.3  & 531M  & 13h   & \multicolumn{1}{r}{1$\times$RTX4090-24G} \\
    MEIC-DT(\textbf{ours}) & DeBERTa$_\text{large}$ & 90.5  & 84.4  & 80.2  & \textbf{85.1}  & 479M  & 14h   & \multicolumn{1}{r}{1$\times$RTX4090-24G} \\
    \hline
    \rowcolor{mypink}\multicolumn{9}{l}{\textbf{Generative methods}} \\
    ASP~\cite{liu-etal-2022-autoregressive}   & FLAN-T5$_\text{XXL}$ & 87.2  & 81.7  & 78.6  & 82.5  & 11B   & 45h   & \multicolumn{1}{r}{6$\times$A100-80G} \\
    Link-Append~\cite{bohnet-etal-2023-coreference} & mT5$_\text{XXL}$ & 87.8  & 82.6  & 79.5  & \textbf{83.3}  & 13B   & 48h   & \multicolumn{1}{r}{128$\times$TPUv4-32G} \\
    seq2seq~\cite{zhang-etal-2023-seq2seq} & T0$_\text{XXL}$ & 87.6  & 82.4  & 79.5  & 83.2  & 11B   & -     & \multicolumn{1}{r}{8$\times$A100-80G} \\
    \hline
    \rowcolor{mypink}\multicolumn{9}{l}{\textbf{Large Language Models}} \\
    \multirow{2}[0]{*}{InstructGPT~\cite{le2023large}} & \multirow{2}[0]{*}{API} & 70.4  & 58.4  & 51.7  & 60.1  & -     & -     & \multicolumn{1}{c}{-} \\
          &       & 89.2  & 79.4  & 73.7  & 80.8$^\star$  & -     & -     & \multicolumn{1}{c}{-} \\
    \hdashline
    \multirow{2}[0]{*}{GPT-3.5-turbo~\cite{le2023large}} & \multirow{2}[0]{*}{API} & 66.9  & 55.5  & 46.5  & 56.3  & -     & -     & \multicolumn{1}{c}{-} \\
          &       & 86.2  & 79.3  & 68.3  & 77.9$^\star$  & -     & -     & \multicolumn{1}{c}{-} \\
    \hdashline
    \multirow{2}[1]{*}{GPT-4~\cite{le2023large}} & \multirow{2}[1]{*}{API} & 73.7  & 62.7  & 52.3  & 62.9  & -     & -     & \multicolumn{1}{c}{-} \\
          &       & 93.7  & 88.8  & 82.8  & \textbf{88.4}$^\star$  & -     & -     & \multicolumn{1}{c}{-} \\
    \hline
    \rowcolor{mypink}\multicolumn{9}{l}{\textbf{Supervised neural methods+Large Language Models}} \\
    ImCoref-CeCo$_\text{gpt4o}$~\cite{luo2025imcoref} & DeBERTa$_\text{large}$+API & 91.2  & 84.9  & 81.8  & \textbf{86.0}  & 531M  & 13h   & 1$\times$RTX4090-24G \\
    \bottomrule
    \end{tabular}%
    }
    \caption{Results of different methods on OntoNotes. We report the MUC, B$^3$, and CEAF$_{\phi_4}$ F1 scores (\%), their average (Avg. F1), base encoders, model parameters (Params), and training time/hardware. ($\diamond$) indicates models trained with additional resources; ($\star$) signifies clustering was performed using gold mentions. For the LLMs utilized, we access them remotely via their API interfaces, uniformly setting the temperature parameter to $0$.}
  \label{tab:Per_com_ontonotes}%
\end{table*}%

\begin{table}[htbp]
  \centering
  \resizebox{1.0\columnwidth}{!}{
    \begin{tabular}{lccc|c}
    \toprule
    \multicolumn{1}{c}{Methods} & MUC   & B$^3$   & CEAF$_{\phi_4}$ &\textcolor[rgb]{0.8,0,0}{\textbf{Avg.F1}} \\
    \hline
    longdoc~\cite{toshniwal2021generalization} & -     & -     & -     & 60.1  \\
    Dual-cache~\cite{guo-etal-2023-dual} & 72.1  & 62.1  & 54.7  & 63.0  \\
    LingMess~\cite{otmazgin2022lingmess} & -     & -     & -     & 62.6  \\
    Maverick$_{\text{mes}}$~\cite{martinelli-etal-2024-maverick} & 81.5  & 65.4  & 53.5  & 66.8  \\
    Maverick$_{\text{incr}}$~\cite{martinelli-etal-2024-maverick} & 81.2  & 65.5  & 53.7  & 66.8  \\
    ImCoref$_{\text{mes}}$~\cite{luo2025imcoref} & 80.6  & 66.7  & 55.6  & 67.6  \\
    MEIC-DT(\textbf{ours}) & 80.2  & 68.4  & 69.4 & \textbf{72.6}  \\
    \hline
    InstructGPT~\cite{le2023large} & -     & -     & -     & \textbf{72.9}  \\
    GPT-3.5-turbo~\cite{le2023large} & -     & -     & -     & 70.8  \\
    \hline
    ImCoref-CeS$_{\text{gpt4}}$~\cite{luo2025imcoref} & 83.6  & 69.8  & 66.1  & \textbf{73.2}  \\
    \bottomrule
    \end{tabular}%
    }
    \caption{Performance comparison (\%) of different methods on WikiCoref.}
  \label{tab:Per_com_wikiCoref}%
  \vspace*{-1ex}
\end{table}%

\end{document}